\documentclass[aps,prd,twocolumn,groupedaddress,amssymb,eqsecnum,showpacs,
nofootinbib]{revtex4}
\usepackage{graphicx}
\usepackage{bm}

\begin{document}

\preprint{UMHEP-460}
\preprint{CMU-HEP-04-10}

\title{Renormalization of initial conditions and the trans-Planckian problem of inflation}

\author{Hael Collins}
\email{hael@physics.umass.edu}
\affiliation{Department of Physics, University of Massachusetts, 
Amherst MA 01003}
\author{R.~Holman}
\email{rh4a@andrew.cmu.edu}
\affiliation{Department of Physics, Carnegie Mellon University, 
Pittsburgh PA 15213}

\date{\today}

\begin{abstract}
Understanding how a field theory propagates the information contained in a given initial state is essential for quantifying the sensitivity of the cosmic microwave background to physics above the Hubble scale during inflation.  Here we examine the renormalization of a scalar theory with nontrivial initial conditions in the simpler setting of flat space.  The renormalization of the bulk theory proceeds exactly as for the standard vacuum state.  However, the short distance features of the initial conditions can introduce new divergences which are confined to the surface on which the initial conditions are imposed.  We show how the addition of boundary counterterms removes these divergences and induces a renormalization group flow in the space of initial conditions.
\end{abstract}


\maketitle

\section{Introduction}
\label{intro}

A fundamental principle of field theory is that we do not need to know the details of a particular theory at arbitrarily short distances to be able to use it to make physical predictions.  This ignorance is permissible not because the corrections from high energies are small; in fact they usually are infinitely large. Instead, our ignorance of the short distance behavior of the theory can be absorbed into the redefinition of the parameters describing the theory. The predictions can then be expressed in terms of a finite number of parameters associated with a set of local operators.  In this renormalized theory, higher order corrections, which are usually defined as power series in a small rescaled coupling, remain small.  What we have lost in this process is the idea of fixed, constant parameters; coupling ``constants'' now depend upon the scale at which they have been defined.

The behavior of a field theory in the early universe appears to violate this principle of decoupling---at least at a first glance.  A distinctive feature of inflation \cite{textbooks} is the superluminal stretching of length scales which allows quantum field fluctuations during the inflationary phase to induce the metric perturbations that seed the large scale structure of the universe. The observed spectrum of acoustic peaks in the cosmic microwave background (CMB) and, even more strikingly, the detection of anticorrelations in the polarization and temperature anisotropies at superhorizon scales are both completely consistent with the predictions of inflation \cite{wmap}.  How much expansion occurred during inflation depends on the expansion rate and its duration, but with more than the 60--70 $e$-folds usually demanded of inflation, the scales associated with the large-scale structure we see today would have had their origin in fluctuations at sub-Planckian scales during the early universe.  Most models for inflation typically produce significantly more expansion than this minimum requirement.  This apparent violation of decoupling has been called the ``trans-Planckian problem'' of inflation \cite{brandenberger}, but it can also be regarded as an opportunity, since it suggests that physics that is important at length scales much smaller than those accessible to accelerator experiments could have left its imprint on the CMB.

Because of this unique opportunity, much effort \cite{gary,kaloper,transplanck,cliff,fate,neweasther} has been devoted recently to determining under what conditions such effects could be seen, usually within a particular framework for the near-Planck scale physics that determines the state of the inflaton.  The expansion rate during inflation defines a natural scale, $H$, the Hubble scale, and if the new physics above this scale has a mass $M$ associated with it, then the ``trans-Planckian'' signal is generally found to be suppressed by $H/M$, or more, relative to the vacuum result.  Here the vacuum corresponds to the state that is invariant under the symmetries of the background space-time and that matches with the usual idea for the Minkowski space vacuum at distances much shorter than the curvature of the background, $1/H$.  While these approaches have been very illuminating since they have provided a clear estimate for the expected size of the trans-Planckian signal in the CMB, they have generally been rather {\it ad hoc\/} in that they assume some particular feature for the physics near the Planck scale or they chose a particular initial state.  In this sense, they do not really address the underlying trans-Planckian problem---why we should expect to observe a CMB spectrum that essentially agrees with that produced by assuming that the universe is in its vacuum state at all scales, without any {\it a priori\/} assumptions about physics near the Planck scale.

The setting for a field theory in the early universe is somewhat different than that generally assumed in an $S$-matrix calculation \cite{witten}.  The field begins in some state specified at an initial time, $t_0$, which does not need to correspond to an asymptotically well-behaved state in the infinite past or which was an energy eigenstate of the free Hamiltonian.  Depending upon the space-time geometry, a globally conserved energy may not even exist.  This initial state can have short distance features imprinted upon it either from a preceding phase where heavy fields were excited or at length scales shorter than the Planck length, where gravitational effects can become strong.

What is therefore needed is an extension of the ideas of effective field theory \cite{eft} for describing the features of an arbitrary initial state.  In this article we describe how the renormalization of this short distance information in the initial state proceeds in Minkowski space.  An initial state can be broadly classified as renormalizable or nonrenormalizable by how its behavior differs from the vacuum state at short distances.  The new divergences associated with the short distance features of the initial state are cancelled by counterterms localized at the surface where the initial state is defined.  Not surprisingly, the renormalizable initial states are associated with relevant or marginal operators on the initial surface while the nonrenormalizable initial states require irrelevant boundary counterterms.

To begin, we must understand what constitutes a renormalizable initial state---where the divergences in the bare theory occur and how they can be removed by the appropriate counterterms which are consistent with the symmetries left unbroken by the state.  An important aspect is the statement of a renormalization condition for this setting, since it is through this condition that the renormalized theory acquires its dependence on the renormalization scale.  The Callan-Symanzik equation then determines the running of the usual couplings and field rescalings as well as the scale dependence of the effects associated with the initial condition.

The second important aspect of an effective treatment of the initial state is characterizing and understanding the role of nonrenormalizable initial conditions.  In the modern view of field theory, nonrenormalizable operators are not unphysical but are merely a sign that a theory is not intended to be valid to arbitrarily high energies.  The coefficient of a nonrenormalizable operator of dimension $n>4$ is of the order $1/M^{n-4}$, where $M$ is a mass scale.  If we study phenomena at an energy $\Lambda$, then the effects of the nonrenormalizable operators will be suppressed by powers $\Lambda/M$.  As long as $\Lambda\ll M$ and we do not demand an arbitrary accuracy of our prediction, we only need to consider a finite set of operators.  As $\Lambda\to M$, the effective description breaks down.  But once we are able to probe such energies, we should see the dynamics that gave rise to the higher dimension operators of our lower energy effective theory and that description is replaced by another effective theory appropriate for the scale $\Lambda\sim M$.

Similarly, nonrenormalizable initial conditions naturally introduce a mass scale $M$.  To make a prediction at a length $1/\Delta$ sufficiently larger than $1/M$ and to specified accuracy $\delta$---and in the measurement of the CMB we shall probably always have a larger error than we would like---then we need only that finite set of parameters describing the initial state suppressed by no more than $n$ powers of $\Delta/M$ where $n$ is defined by $(\Delta/M)^n \sim \delta$.  There are additional parameters associated with the still finer details of the initial state, but these are increasing irrelevant, being suppressed by further powers of $\Delta/M$ and are not needed in practice until we can make more precise measurements (decreasing $\delta$) or are able to probe the shorter distance features directly (increasing $\Delta$).  What has happened is that we have replaced an assumption about the details of the Planck scale physics setting the initial state with a small set of parameters needed to describe how the state differs from the vacuum at short distances and which is applicable for any short distance completion of the theory.

Despite the power and the generality of the effective field theory perspective, this approach has not been widely applied to inflation.  Ref.~\cite{kaloper} derived the expected size of the corrections to the CMB power spectrum from higher order operators, but still worked in the standard vacuum state.  One of the earliest attempts to model initial state effects in the spirit of effective field theory integrated out the dynamics of a heavy field to learn of its imprint on the CMB \cite{cliff}; but the first attempt for a general effective description of initial state effects has only appeared quite recently \cite{schalm,schalm2}. 

The essential feature of the renormalization of an initial state to understand is how to control the infinities that arise from having short distance features in this state that differ from the standard vacuum.  It is important to note that for a Robertson-Walker universe the vacuum corresponds to the state that matches with the Minkowski space vacuum at distances sufficiently small that the background curvature is not noticeable.  Since the curvature is unimportant where the renormalization is needed, the appropriate setting in which to begin is Minkowski space.  It is also simpler in flat space to examine the structure of a state and its renormalization analytically.  As we consider a completely general initial state, we shall use the Schwinger-Keldysh formalism \cite{schwinger,keldysh,kt} to define the time-evolution of a general Green's function in this setting.  

Once we have understood the initial state renormalization in flat space, we shall have the necessary foundation for proceeding to a general Robertson-Walker space-time.  The expansion of the background affects how and when we should choose an initial state.  The expansion rate $H$, from an effective theory perspective, just sets the appropriate scale with respect to which we define the infrared (IR) and ultraviolet (UV) details of the initial state.  But the continual blueshifting that occurs as we look further back during inflation means that we must also choose an appropriate initial time at which to set the initial conditions defining the state \cite{schalm2}, which is naturally and readily accomplished in the Schwinger-Keldysh picture.  Both of these details will be addressed in \cite{initfrw}.

The next section begins with a description of how an initial condition alters the structure of the propagator.  The need for the consistency of the propagator with the state is quite familiar from studies of interacting theories \cite{einhorn,lowe,taming} for the $\alpha$-vacua of de Sitter space \cite{alpha}.  One of the new features here is a part of the propagator that depends on the initial state, which gives rise to divergences not present for the vacuum state.  Section \ref{rencon} describes how to renormalize the theory in this setting, by establishing an appropriate renormalization condition and showing that we obtain the usual running for the state-independent part of the theory.  Section \ref{bndren} then shows that all of the initial-state-dependent divergences are localized on the boundary where the initial condition was imposed.  We then remove these divergences with boundary counterterms and derive their running dependence on the renormalization scale.  We describe nonrenormalizable initial conditions in Sec.~\ref{nonrenorm} using the language of effective field theory.  Section \ref{conclude} concludes with comments on the extension of this approach to Robertson-Walker universes and to the problem of backreaction.  Two more detailed points of our discussion are presented in the appendices.  The first provides a brief introduction to the Schwinger-Keldysh formalism while the second shows how a resummation removes some spurious logarithmic divergences at the initial boundary which do not affect the renormalization.

\section{Boundary conditions in flat space}
\label{flat}

Consider a free massive scalar field propagating in a flat space-time,
\begin{equation}
S = \int d^4x\, \left[ 
{\textstyle{1\over 2}} \partial_\mu\varphi \partial^\mu\varphi 
- {\textstyle{1\over 2}} m^2 \varphi^2 \right] . 
\label{freeaction}
\end{equation}
The usual expansion of the field in creation and annihilation operators with respect to the vacuum is 
\begin{equation}
\varphi(t,\vec x) = \int {d^3\vec k\over (2\pi)^3\sqrt{2\omega_k}}\, \left[ 
e^{-i\omega_k t} e^{i\vec k\cdot \vec x} a_{\vec k}
+ e^{i\omega_k t} e^{-i\vec k\cdot \vec x} a_{\vec k}^\dagger \right] ,
\label{vacuummodes}
\end{equation}
where the frequency is $\omega_k = \sqrt{k^2+m^2}$.  

We shall generalize this mode expansion to the situation where the behavior of the field is specified along an initial space-like surface, $t=t_0$, so it is instructive to recall the origin of the form of the modes.  Since the initial surface is spatially flat, the spatial modes are still plane waves but we leave the time-dependent part unspecified, 
\begin{equation}
\varphi(t,\vec x) = \int {d^3\vec k\over(2\pi)^3}\, \left[ 
U_k(t) e^{i\vec k\cdot \vec x} a_{\vec k}
+ U_k(t)^* e^{-i\vec k\cdot \vec x} a_{\vec k}^\dagger \right] . 
\label{genmodes}
\end{equation}
The time-dependent part of the mode functions, $U_k(t)$, is the solution to the Klein-Gordon equation, 
\begin{equation}
U_k(t) = c_k e^{-i\omega_k t} + d_k e^{i\omega_k t} .
\label{gensoln}
\end{equation}
One of the constants of integration is already fixed by the equal time commutator, 
\begin{equation}
\bigl[ \pi(t,\vec x), \varphi(t,\vec y) \bigr] = -i \delta^3(\vec x-\vec y) , 
\qquad 
\pi = {\partial\varphi\over\partial t} , 
\label{etc}
\end{equation}
which imposes 
\begin{equation}
U_k\partial_t U_k^* - U_k^*\partial_t U_k = i . 
\label{wronk}
\end{equation}
Applying this Wronskian condition to Eq.~(\ref{gensoln}) yields 
\begin{equation}
|c_k|^2 - |d_k|^2 = {1\over 2\omega_k} . 
\label{wronkongen}
\end{equation}

The second of the constants in Eq.~(\ref{gensoln}) is determined by an additional, physically motivated condition.  The standard condition is to choose the modes to be those associated with the vacuum; selecting only the positive energy states imposes $d_k=0$, which together with the equal time commutation relation establishes the usual form for the modes, 
\begin{equation}
U_k^E(t) = {1\over\sqrt{2\omega_k}} e^{-i\omega_k t} , 
\label{flatmodes}
\end{equation}
up to an arbitrary phase.  The ``$E$'' signals that the modes are those of the vacuum.

Notice that the state annihilated by all of the $a_{\vec k}$,
\begin{equation}
a_{\vec k}\, |0\rangle = 0 , 
\label{flatvac}
\end{equation}
is a vacuum state both in the sense of being invariant under the generators of the Poincar\' e group as well as being the lowest energy eigenstate with respect to the globally conserved Hamiltonian associated with the Killing vector ${\partial\over\partial t}$.  In a de Sitter or a Robertson-Walker background, a globally defined time-like Killing vector may not exist and so for these space-times we can only define a vacuum in terms of its symmetries.

For a more general initial state, we fix the second constant of integration in Eq.~(\ref{gensoln}) with an initial condition on the modes.  This approach is especially useful in the early universe where the rapid expansion does not guarantee the existence of a noninteracting state in the asymptotic past.  For simplicity we restrict ourselves to a first-order constraint that is linear in the mode functions,
\begin{equation}
{\partial\over\partial t} U_k(t) \biggr|_{t=t_0} = - i\varpi_k U_k(t_0) . 
\label{BDflat}
\end{equation}
The particular boundary condition with $\varpi_k=\omega_k$ restores the full Poincar\' e invariance of the state, but other values for $\varpi_k$ break the symmetry.  Applying this condition to the general mode function in Eq.~(\ref{gensoln}) with Eq.~(\ref{wronkongen}) yields 
\begin{equation}
U_k(t) = {|\omega_k+\varpi_k|\over 2\omega_k\sqrt{2{\rm Re} \varpi_k}} \left[ 
e^{-i\omega_k t} + {\omega_k-\varpi_k\over\omega_k+\varpi_k} e^{i\omega_k (t-2t_0)}
\right] .
\label{BDmodesI}
\end{equation}

We can rewrite these mode functions in a form that emphasizes the similarity of these modes to the mode functions used for the $\alpha$-states of de Sitter space \cite{alpha}.  Define 
\begin{equation}
e^{\alpha_k} \equiv {\omega_k-\varpi_k\over\omega_k+\varpi_k}
\label{ealphdef}
\end{equation}
along with an {\it image time\/}, 
\begin{equation}
t_{\scriptscriptstyle I} \equiv 2t_0 - t ,
\label{timagedef}
\end{equation}
so that 
\begin{equation}
U_k^\alpha(t) = {N_k\over\sqrt{2\omega_k}} \left[ 
e^{-i\omega_k t} + e^{\alpha_k} e^{-i\omega_k t_{\scriptscriptstyle I}}
\right]
\label{BDmodes}
\end{equation}
with 
\begin{equation}
N_k \equiv \left[ 1 - e^{\alpha_k+\alpha_k^*} \right]^{-1/2}
= {|\omega_k+\varpi_k|\over 2\sqrt{\omega_k\, {\rm Re} \varpi_k}} .
\label{Nalphdef}
\end{equation}
In terms of the Poincar\' e invariant modes,
\begin{equation}
U_k^E(t) = {e^{-i\omega_k t}\over\sqrt{2\omega_k}} ,
\label{PImodes}
\end{equation}
the modes satisfying the boundary condition are then 
\begin{equation}
U_k^\alpha(t) 
= N_k \left[ U_k^E(t) 
+ e^{\alpha_k} U_k^E(t_{\scriptscriptstyle I})
\right] . 
\label{BDmodesalph}
\end{equation}
The ``$\alpha$'' indicates the mode functions, and later the Green's functions, that are consistent with the boundary condition of Eq.~(\ref{BDflat}).

Since the modes can be written in terms of the flat space modes, the creation and annihilation operators associated with the initial state can be correspondingly written as a Bogoliubov transformation of the flat space operators,
\begin{equation}
a_{\vec k}^{\alpha_k}  
= N_k \left[ a_{\vec k}
+ e^{\alpha_k^*} e^{2i\omega_k t_0} a_{-\vec k}^\dagger \right] . 
\label{bogolubov}
\end{equation}
Let us write the state that is annihilated by the $a_{\vec k}^{\alpha_k}$ operators at $t=t_0$ as 
\begin{equation}
a_{\vec k}^{\alpha_k}\, |\alpha_k(t_0)\rangle = 0 . 
\label{bndstate}
\end{equation}
For simplicity, we shall often write the initial state without its argument,
\begin{equation}
|\alpha_k\rangle \equiv |\alpha_k(t_0)\rangle . 
\label{alphanot0}
\end{equation}
In the interaction picture, which we follow here, the time-evolution of operators in the theory is given by free Hamiltonian while the interacting part of the Hamiltonian generates the evolution of the states, 
\begin{equation}
|\alpha_k(t)\rangle = U(t,t_0)\, |\alpha_k(t_0)\rangle , 
\label{UIdef}
\end{equation}
where $U(t,t_0)$ is the operator solving Dyson's equation and is written in Eq.~(\ref{Udyson}).  But before analyzing an interacting theory we must first establish a description for the propagation of a field in the region $t>t_0$ which respects the initial constraint, Eq.~(\ref{BDflat}).

\subsection{Propagation}

The propagator must also be consistent with the boundary conditions, which implies some additional structure beyond that of the usual, Poincar\'e invariant Feynman propagator, 
\begin{eqnarray}
-i G_F^E(x,x') 
&=& \Theta(t-t') \langle 0 | \varphi(x)\varphi(x') | 0 \rangle
\nonumber \\
&&
+ \Theta(t'-t) \langle 0 | \varphi(x')\varphi(x) | 0 \rangle
\nonumber \\
&=& \int {d^4k\over (2\pi)^4}\, 
{ie^{-ik\cdot(x-x')}\over k^2-m^2+i\epsilon} , 
\label{feynman}
\end{eqnarray}
since the derivatives in the boundary conditions can also act on the $\Theta$-functions.  As with the fields, it is simplest to describe the constraint on the propagator in terms of its spatial Fourier transform, 
\begin{equation}
G_F^{\alpha_k}(x,x') = \int {d^3\vec k\over (2\pi)^3}\, 
e^{i\vec k\cdot (\vec x-\vec x')} G_k^{\alpha_k}(t,t') . 
\label{Aprop}
\end{equation}
If we impose the conditions, 
\begin{eqnarray}
\left[ \partial_t - i\varpi_k \right]_{{t=t_0\atop t'>t_0}} 
G_k^{\alpha_k}(t,t') &=& 0 
\nonumber \\
\left[ \partial_{t'} - i\varpi_k \right]_{{t'=t_0\atop t>t_0}} 
G_k^{\alpha_k}(t,t') &=& 0 , 
\label{propBC}
\end{eqnarray}
then the propagator in the region $t,t'>t_0$ is 
\begin{equation}
G_k^{\alpha_k}(t,t') = A_k \left[ G_k^E(t,t') 
+ e^{\alpha_k}\, G_k^E(t_{\scriptscriptstyle I},t') \right] . 
\label{ApropasEAk}
\end{equation}
where $G_k^E(t,t')$ is the momentum representation of the standard vacuum propagator of Eq.~(\ref{feynman}).  The same propagator is used in \cite{schalm}.

In this form, the propagator appears to contain two sources, at $x=x'$ and at $x_{\scriptscriptstyle I}=x'$.  The first term corresponds to the effect of a point particle and the standard normalization of the residue at the physical pole at $k^2=m^2$ fixes $A_k=1$, 
\begin{equation}
G_k^{\alpha_k}(t,t') = G_k^E(t,t') 
+ e^{\alpha_k}\, G_k^E(t_{\scriptscriptstyle I},t') . 
\label{ApropasE}
\end{equation}
The second term represents a source in the unphysical region $t<t_0$; through it we have exchanged a constraint on the boundary with a bulk effect.  The apparent nonlocality in the correlated motion of the particle with its image encodes the fact that we have specified the value of the field over an entire space-like hypersurface.

The propagator in Eq.~(\ref{ApropasE}) also follows from a generalized construction for the time-ordering operator.  Consider a time-ordering which includes the image time defined by 
\begin{eqnarray}
&&\!\!\!\!\!\!\!\!\!\!\!\!\!\!\!\!\!\!\!\!
T_{\alpha_k}\bigl( \varphi(x)\varphi(x') \bigr) 
\label{Talph} \\
&=& \hat\Theta_1(x,x')\, \varphi(x)\varphi(x') 
+ \hat\Theta_2(x,x')\, \varphi(x')\varphi(x) 
\nonumber \\
&& + \hat\Theta_3(x,x')\, \varphi(x_{\scriptscriptstyle I})\varphi(x') 
+ \hat\Theta_4(x,x')\, \varphi(x')\varphi(x_{\scriptscriptstyle I})
\nonumber
\end{eqnarray}
where 
\begin{eqnarray}
\hat\Theta_1(x,x') &=& \hat A\, \Theta(t-t') 
+ {e^{\alpha_k}(e^{\alpha_k}\hat A - \hat B)\over 1 - e^{2\alpha_k}}
\nonumber \\
\hat\Theta_2(x,x') &=& \hat A\, \Theta(t'-t) 
+ {e^{\alpha_k^*}(e^{\alpha_k^*}\hat A - \hat B)\over 1 - e^{2\alpha_k^*}}
\nonumber \\
\hat\Theta_3(x,x') &=& \hat B\, \Theta(t_{\scriptscriptstyle I}-t') 
+ {e^{\alpha_k}(e^{\alpha_k}\hat B - \hat A)\over 1 - e^{2\alpha_k}}
\nonumber \\
\hat\Theta_4(x,x') &=& \hat B\, \Theta(t'-t_{\scriptscriptstyle I}) 
+ {e^{\alpha_k^*}(e^{\alpha_k^*}\hat B - \hat A)\over 1 - e^{2\alpha_k^*}} . 
\nonumber 
\label{Thetas}
\end{eqnarray}
The propagator in Eq.~(\ref{ApropasE}) is obtained when 
\begin{equation}
\hat A = 1 
\qquad
\hat B = \widehat{e^{\alpha_k}} 
\label{ABfix}
\end{equation}
by evaluating this time-ordering in an $|\alpha_k\rangle$ states, 
\begin{equation}
-i G_F^{\alpha_k}(x,x') =
\langle \alpha_k | T_{\alpha_k}\bigl( \varphi(x)\varphi(x') \bigr) | \alpha_k \rangle . 
\label{Talphprop} 
\end{equation}
Here, the operators are defined in terms of their action on the Fourier components of the field, as for example, 
\begin{equation}
\widehat{e^{\alpha_k}}\, f(t,\vec x) = \int {d^3\vec k\over (2\pi)^3}\, 
e^{i\vec k\cdot \vec x} e^{\alpha_k} f_{\vec k}(t) . 
\label{ealphaOP} 
\end{equation}
Note that in the particular case where $e^{\alpha_k}$ is real, which implies that $\varpi_k$ is also real, the time ordering becomes especially simple,
\begin{eqnarray}
&&\!\!\!\!\!\!\!\!
T_{\alpha_k}\bigl( \varphi(x)\varphi(x') \bigr) 
\nonumber \\
&&= \Theta(t-t')\, \varphi(x)\varphi(x') 
+ \Theta(t'-t)\, \varphi(x')\varphi(x) 
\nonumber \\
&&\quad 
- \widehat{e^{\alpha_k}}\, 
\bigl[ 
\Theta(t'-t_{\scriptscriptstyle I}) \, \varphi(x_{\scriptscriptstyle I})\varphi(x') 
\nonumber \\
&&\quad\qquad\ 
+ \Theta(t_{\scriptscriptstyle I}-t')\, \varphi(x')\varphi(x_{\scriptscriptstyle I}) \bigr]. 
\label{Talpheasy} 
\end{eqnarray}
The time-ordering in the image $\Theta$-functions is the opposite that of the fields---forward propagation of the physical particle corresponds to the backward propagation of its image.

We would like to show the relation between propagator derived above and that used in \cite{schalm}; both propagators are essentially the same---and agree within the physical region $t,t'>t_0$---but the statement of the boundary conditions differs slightly from that which we have used in Eq.~(\ref{BDflat}).  In \cite{schalm} the propagator\footnote{See for example Eq.~(2.17) of \cite{schalm}.} is written in the form, 
\begin{equation}
G_F^{\alpha_k}(x,x') = \int {d^4k\over (2\pi)^4}\, 
{e^{-ik\cdot (x-x')} + e^{\alpha_k(k_0)} e^{-ik\cdot (x_{\scriptscriptstyle I}-x')}\over k^2 - m^2 + i\epsilon}
\label{dprop}
\end{equation}
where $x^\mu_{\scriptscriptstyle I}=(t_{\scriptscriptstyle I},\vec x)$.  Unlike the single-source propagator used in the Poincar\' e-invariant vacuum state of Eq.~(\ref{feynman}), this propagator contains an additional dependence on $k_0$ through the prefactor of the image term, 
\begin{equation}
e^{\alpha_k(k_0)} = {k_0+\varpi_k\over k_0 - \varpi_k}
\label{badpole}
\end{equation}
This coefficient leads to a spurious new pole term, at $k_0=\varpi_k$, which does not actually affect the propagator in the region $t\ge t_0$ provided we assume that $\varpi_k$ has a small negative imaginary part.  Therefore, although Eq.~(\ref{dprop}) more closely resembles the standard expression for the flat space propagator integrated over $k_0$ as well, we shall continue to write the propagator in its explicitly time-ordered form with an integral over only the spatial wave vector.

Let us take the Fourier transform of the propagator in Eq.~(\ref{dprop}), 
\begin{eqnarray}
G_F^{\alpha_k}(x,x') 
&=& \int {d^3\vec k\over (2\pi)^3}\, 
e^{i\vec k\cdot (\vec x-\vec x')} G_k^{\alpha_k}(t,t').
\label{dpropFT}
\end{eqnarray}
Inserting $\Theta$-functions appropriately, we obtain
\begin{eqnarray}
&&\!\!\!\!\!\!\!\!\!\!\!\! G_k^{\alpha_k}(t,t') 
\label{dpropTH}  \\
&=& \Theta(t-t') \oint_{\rm lower} {dk_0\over 2\pi}\, 
{e^{-ik_0(t-t')} \over k_0^2 - (\omega_k^2 - i\epsilon)}
\nonumber \\
&& 
+ \Theta(t'-t) \oint_{\rm upper} {dk_0\over 2\pi}\, 
{e^{ik\cdot (t'-t)} \over k_0^2 - (\omega_k^2 - i\epsilon)}
\nonumber \\
&& 
+ \Theta(t_{\scriptscriptstyle I}-t') \oint_{\rm lower} {dk_0\over 2\pi}\, 
{k_0+\varpi_k\over k_0 - \varpi_k}
{e^{-ik\cdot (t_{\scriptscriptstyle I}-t')}\over k_0^2 - (\omega_k^2 - i\epsilon)}
\nonumber \\
&& 
+ \Theta(t'-t_{\scriptscriptstyle I}) \oint_{\rm upper} {dk_0\over 2\pi}\, 
{k_0+\varpi_k\over k_0 - \varpi_k}
{e^{ik\cdot (t'-t_{\scriptscriptstyle I}-t')}\over k_0^2 - (\omega_k^2 - i\epsilon)} , 
\nonumber 
\end{eqnarray}
which allows us to integrate each of the terms by closing the contour in the upper or the lower half plane as indicated.  Note that although we have inserted $\Theta$-functions for the image time as well, $\Theta(t_{\scriptscriptstyle I}-t')=0$ and $\Theta(t'-t_{\scriptscriptstyle I})=1$ since we are restricted to $t,t'>t_0$.  Therefore, the third term always vanishes in the physical region.  Assuming that ${\rm Im}\, \varpi_k < 0$, the extra factor in the fourth term does not produce any new poles so that the contour integral gives only the usual result, 
\begin{eqnarray}
G_k^{\alpha_k}(t,t') 
&\!=\!& \Theta(t-t') {i\over 2\omega_k} e^{-i\omega_k(t-t')} 
\nonumber \\
&&
+ \Theta(t'-t) {i\over 2\omega_k} e^{-i\omega_k\cdot (t'-t)} 
\nonumber \\
&& 
+ \Theta(t_{\scriptscriptstyle I}-t') {i e^{\alpha_k}\over 2\omega_k} e^{-i\omega_k\cdot (t_{\scriptscriptstyle I}-t')}
\nonumber \\
&& 
+ \Theta(t'-t_{\scriptscriptstyle I}) {i e^{\alpha_k}\over 2\omega_k} e^{-i\omega_k\cdot (t'-t_{\scriptscriptstyle I})} , \qquad
\label{dpropTHOI} 
\end{eqnarray}
It is important to note that the third term in this equation did not result from the third term in Eq.~(\ref{dpropTH}).  Since both vanish for $t,t'>t_0$, we have formally included the appropriate counterpart of the final image term to produce an {\it image propagator\/} when the full propagator is written in its time-ordered form.  This result establishes that the propagators in Eqs.~(\ref{ApropasE}) and (\ref{dprop}) agree in the physical region.

\subsection{Generating functionals}

To generate a propagator with two sources, the free field generating functional should include a current $J(x)$ that couples to the field simultaneously at the point $x$ and its image $x_{\scriptscriptstyle I}$.  This apparent nonlocality in the bulk physics encodes the effect of having imposed a boundary condition upon the spatial hypersurface, $t=t_0$.  The free field generating functional is thus 
\begin{equation}
W_0^{\alpha}[J] = \int {\cal D}\varphi\, e^{i\int d^4x\, 
\left[ {\cal L}_0
+ (\hat a(x) \varphi(x) + \hat b(x_{\scriptscriptstyle I}) \varphi(x_{\scriptscriptstyle I}) ) J(x) \right] }
\label{genfunc0}
\end{equation}
with 
\begin{equation}
{\cal L}_0 = {\textstyle{1\over 2}} \partial_\mu\varphi \partial^\mu\varphi 
- {\textstyle{1\over 2}} m^2 \varphi^2 . 
\label{freeL}
\end{equation}
To obtain the correct propagator we require that the coefficients of the currents should satisfy
\begin{eqnarray}
\hat a &=& {1\over\sqrt{2}} \Bigl[ \hat A + \sqrt{\hat A^2 - \hat B^2} \Bigr]^{1/2} 
\nonumber \\
\hat b &=& {1\over\sqrt{2}} 
{\hat B\over \bigl[ \hat A + \sqrt{\hat A^2 - \hat B^2} \bigr]^{1/2} } 
\label{abdefs}
\end{eqnarray}
which yields the propagator Eq.~(\ref{Talphprop}) when we differentiate with respect to the currents, 
\begin{equation}
-i G_F^{\alpha_k}(x,x') = \left[ -i {\delta\over\delta J(x)} \right]
\left[ -i {\delta\over\delta J(x')} \right] W_0^\alpha[J] \Bigr|_{J=0} . 
\label{propfromW0}
\end{equation}

When the generating functional is generalized to a fully interacting theory, it is important to distinguish the locality of the bulk theory from the nonlocality introduced by the boundary conditions.  The underlying locality of the theory implies that the free Lagrangian remains of the form in Eq.~(\ref{freeL}), but to obtain the correct propagators for the internal lines of a general graph in an interacting theory requires that the interactions should have the form \cite{lowe,taming},
\begin{equation}
{\cal L} = {\textstyle{1\over 2}} \partial_\mu\varphi \partial^\mu\varphi 
- {\textstyle{1\over 2}} m^2 \varphi^2 
- \sum_{n\ge 3} {1\over n!} \lambda_n \tilde\varphi^n 
\label{Lint}
\end{equation}
where
\begin{equation}
\tilde\varphi(x) \equiv \hat a(x) \varphi(x) 
+ \hat b(x_{\scriptscriptstyle I}) \varphi(x_{\scriptscriptstyle I}) . 
\label{tildephi}
\end{equation}
The interacting theory generating function is then 
\begin{equation}
W^{\alpha}[J] = {\cal N}\int {\cal D}\varphi\, e^{i\int d^4x\, 
\left[ {\cal L}_0(x) - \sum_{n\ge 3} {\lambda_n\over n!} \tilde\varphi^n(x)
+ \tilde\varphi(x) J(x) \right] }   
\label{genfunc}
\end{equation}
with 
\begin{equation}
{1\over{\cal N}} = \int {\cal D}\varphi\, e^{i\int d^4x\, 
\left[ {\cal L}_0(x) - \sum_{n\ge 3} {\lambda_n\over n!} \tilde\varphi^n(x)
\right] } . 
\label{Normdef}
\end{equation}
A more convenient form for the generating functional when calculating perturbative corrections is 
\begin{equation}
W^{\alpha}[J] = e^{- i\int d^4x\, 
\sum_{n\ge 3} {\lambda_n\over n!} \left[-i{\delta\over\delta J(x)} \right]^n } 
W^{\alpha}_0[J] ,  
\label{genfuncD}
\end{equation}
with the free field generating functional of Eq.~(\ref{genfunc0}) rewritten in the form
\begin{equation}
W_0^{\alpha}[J] = e^{(i/2) \int d^4xd^4x'\, J(x) G_F^{\alpha_k}(x,x') J(x')} . 
\label{genfunc0shift}
\end{equation}

\section{Renormalization conditions}
\label{rencon}

In the standard formulation of field theory, the perturbative corrections to a given process often diverge as we sum over more and more of the short distance behavior.  A theory remains predictive since, in the case of renormalizable interactions, it is possible to absorb this divergent behavior into a redefinition of the parameters of the theory.  The physical parameters, each given by the sum of the corresponding infinite bare parameter and its infinite radiative corrections, remain finite.  In this process, the parameters acquire a dependence on the scale at which they are defined.

When we consider a nonstandard boundary condition, we introduce an additional type of renormalization.  In this case, we encounter new divergences related to summing over the short distance features of the initial state.  For the theory to remain predictive, we need a comparable method for absorbing our ignorance of the extreme short distance structure of the state with a corresponding infinite counterterm.  Since these new divergences are features of the initial conditions and not the bulk physics, the new counterterms should be confined to the initial surface.

The physical setting is one in which the field may not necessarily be in its vacuum state, so we shall adopt the methods usually applied in nonequilibrium field theory \cite{neft} to establish the renormalization conditions that determine the scale dependence of the various parameters of the theory.  For the ``bulk'' $3+1$ dimensional physics, these conditions produce the same anomalous dimensions and $\beta$-functions we would have anticipated from the $S$-matrix.  This agreement between the renormalization group running of the bulk properties of the theory obtained for a general initial state and the running obtained using the $S$-matrix is a necessary and natural consequence of the fact that we have not modified the short distance properties of the theory.  At very short distances and away from the boundary, the field is not sensitive to the details of the initial state.  The bulk divergences should therefore be unaltered by the initial conditions.  

This behavior still leaves the possibility of new divergences that are associated with the short distance details of the initial state.  Since any state other than the vacuum state necessarily breaks the underlying Poincar\' e invariance of the background, the counterterms needed to cancel these boundary divergences are consistent only with this broken symmetry.  In fact, as we shall show, the new divergences that arise from the image term in the propagator only appear at $t=t_0$.  Therefore the renormalization of the initial state corresponds to the appearance boundary counterterms that depend on the initial condition imposed and that run with the renormalization scale.

Our ultimate goal \cite{initfrw} is eventually to establish a framework that can be applied to the early universe.  We therefore treat the scalar field in much the same way as if it were an inflaton in an expanding background, dividing it into a classical zero mode $\phi(t)$ which only depends on time and a small fluctuation $\psi(t,\vec x)$, 
\begin{equation}
\varphi(t,\vec x) = \phi(t) + \psi(t,\vec x) . 
\label{zerofluc}
\end{equation}
The simplest renormalization condition---the vanishing of the tadpole \cite{neft,weinberg}---is 
\begin{equation}
\langle \alpha_k(t) |\, \psi^+(x) \, | \alpha_k(t) \rangle = 0 . 
\label{nolin}
\end{equation}
In the interaction picture, the time evolution of the state is determined by the interacting part of the Hamiltonian.  In the Schwinger-Keldysh approach, which determines the time-evolution of the full matrix element starting from a specified initial state, this condition becomes 
\begin{equation}
{\langle \alpha_k |\, T_{\alpha} \bigl( \psi^+(x) 
e^{-i\int_{t_0}^\infty dt\, 
\left[ H_I(\phi,\psi^+) - H_I(\phi,\psi^-) \right]}
\bigr) \, | \alpha_k \rangle \over
\langle \alpha_k |\, T_{\alpha} e^{-i\int_{t_0}^\infty dt\, 
\left[ H_I(\phi,\psi^+) - H_I(\phi,\psi^-) \right]} \, | \alpha_k \rangle }
 = 0 ,
\label{nolinfull}
\end{equation}
where the denominator removes the vacuum to vacuum graphs.  The $+$ and $-$ superscripts refer to the result of time-evolving both the in-state and the out-state of the matrix element, respectively.  Appendix~\ref{SKformal} briefly reviews the Schwinger-Keldysh approach and further defines some of the notation we have used.

We illustrate the appearance of new boundary divergences by studying a scalar field with a simple quartic self-coupling, 
\begin{equation}
{\cal L} = {\textstyle{1\over 2}} \partial_\mu\varphi \partial^\mu\varphi 
- {\textstyle{1\over 2}} m^2 \varphi^2 
- {\textstyle{1\over 24}} \lambda \varphi^4 . 
\label{phi4}
\end{equation}
This Lagrangian describes the bare theory.  The perturbative corrections to a Green's function, such as the one-point function we shall examine, contain divergences which can be absorbed by rescaling the parameters of the theory, 
\begin{equation}
\varphi = Z_3^{1/2}\varphi_R,
\quad 
m^2 = {Z_0\over Z_3} m^2_R, 
\quad 
\lambda = {Z_1\over Z_3^2} \lambda_R .
\label{Zdefs}
\end{equation}
In terms of the renormalized theory, $\{ \varphi_R, m_R, \lambda_R\}$, the perturbative corrections are finite.  Equivalently, we could write the Lagrangian in terms of the renormalized parameters,
\begin{eqnarray}
{\cal L} 
&=& {\textstyle{1\over 2}} \partial_\mu\varphi_R \partial^\mu\varphi_R 
- {\textstyle{1\over 2}} m_R^2 \varphi_R^2 
- {\textstyle{1\over 24}} \lambda_R \varphi_R^4 . 
\nonumber \\
&& 
+ {\textstyle{1\over 2}} (Z_3-1) \partial_\mu\varphi_R \partial^\mu\varphi_R 
- {\textstyle{1\over 2}} (Z_0-1) m_R^2 \varphi_R^2 
\nonumber \\
&&
- {\textstyle{1\over 24}} (Z_1-1) \lambda_R \varphi_R^4 . 
\label{phi4R}
\end{eqnarray}
The terms on the last two lines correspond to the counterterms needed to render the theory finite.

The image parts of the propagator produce further divergences on the initial boundary so the theory also requires an additional renormalization.  These boundary divergences are renormalizable in the sense that they can be removed by a set of relevant or marginal operators localized at $t=t_0$, which are consistent with the symmetries left unbroken by the boundary.  For example, for a scalar theory we can have 
\begin{equation}
S_{t=t_0} = \int_{t=t_0} d^3\vec x\, \left\{ 
{\textstyle{1\over 2}} z_0 \varphi \partial_t \varphi
+ {\textstyle{1\over 2}} z_1 m \varphi^2 
+ {\textstyle{1\over 6}} z_2 \varphi^3 
\right\} . 
\label{surfaction}
\end{equation}
We have included a factor of $m$ in the quadratic term so that all the $z_i$ are dimensionless.  Note that we have used capital $Z_i$'s for the bulk renormalization and lowercase $z_i$'s for the boundary renormalization.  Since we have broken time-translation invariance, operators such as 
\begin{equation}
\varphi\partial_t\varphi = {\textstyle{1\over 2}} \partial_t \varphi^2,\ \ 
\varphi\partial_t^2\varphi,\ \ \partial_t\varphi\partial_t\varphi,\ \ \ldots
\label{tdervs}
\end{equation}
are allowed on the initial surface, although only the first operator is marginal since the field has a mass dimension of one.  The surface is still $O(3)$ invariant, so the first operator with a spatial derivative only appears as the irrelevant, dimension four operator, $\vec\nabla\varphi \cdot \vec\nabla\varphi$.  In our example of a $\varphi^4$ theory, the Lagrangian has an additional $\varphi\leftrightarrow -\varphi$ invariance so we have in this case that $z_2=0$ automatically.  Thus, a renormalizable boundary condition in this example only requires two types of boundary renormalization.  In this section we shall show how to characterize such renormalizable initial conditions and to determine how they run under a renormalization group flow.
\begin{figure}[!tbp]
\includegraphics{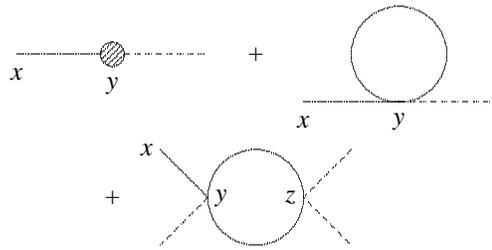}
\caption{The leading contributions to the running of the running of the mass $m$ and the coupling $\lambda$ in a $\varphi^4$ theory.  The solid lines represent propagating $\psi$ fields while the dashed lines correspond to the zero mode $\phi$.
\label{leadloops}}
\end{figure}

\begin{figure}[!tbp]
\includegraphics{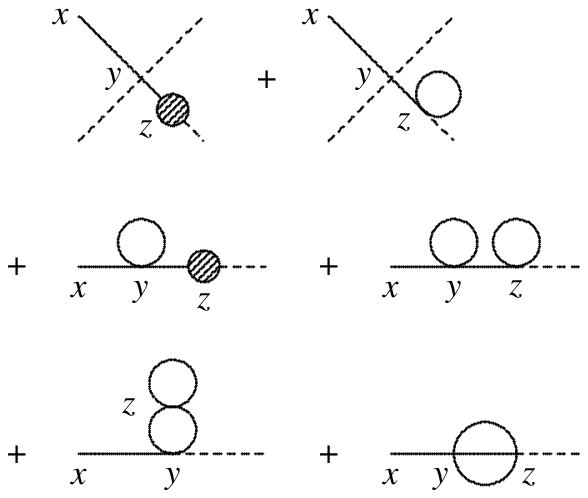}
\caption{Further graphs obtained by expanding the exponential in Eq.~(\ref{nolinfull}) to second order.  The last of these graphs contains the leading nontrivial correction to the wavefunction renormalization.
\label{moreloops}}
\end{figure}

\begin{figure}[!tbp]
\includegraphics{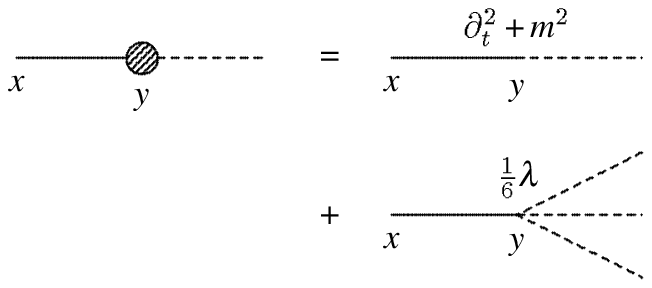}
\caption{The shaded blob corresponds to the following two graphs.  The time derivatives act on the classical $\phi(t)$ field.
\label{opphi}}
\end{figure}

Because of its coupling to the zero mode, the vanishing of the one-point Green's function for the fluctuation contains much information.  From the perspective of $\psi^\pm$, the interacting part of the Hamiltonian is 
\begin{eqnarray}
H_I(\phi,\psi^\pm) 
&=& \int d^3\vec y\, \Bigl[ 
\psi^\pm (\ddot\phi + m^2\phi + {\textstyle{1\over 6}} \lambda \phi^3)
\nonumber \\
&&\qquad
+ {\textstyle{1\over 4}}\lambda \phi^2 {\psi^{\pm}}^2 
+ {\textstyle{1\over 6}}\lambda \phi {\psi^{\pm}}^3 
+ {\textstyle{1\over 24}}\lambda {\psi^{\pm}}^4 \Bigr] . 
\nonumber \\
&&
\label{Hint}
\end{eqnarray}
For simplicity, we shall treat the $\phi^2{\psi^\pm}^2$ term as an interaction rather than as an effective mass term; this treatment is consistent when $\lambda\phi^2\ll m^2$.  More generally, we can resum the effects of this term as has been done in Appendix~\ref{logresum}.  For this interaction, the connected part of the expectation value of $\psi$, expanded to second order in $H_I$, yields
\begin{widetext}
\begin{eqnarray}
0 &=& \langle \alpha_k(t_f) | \psi^+(x) | \alpha_k(t_f) \rangle
\nonumber \\
&=& - \int_{t_0}^{t_f} dt \int d^3\vec y\, 
\left[ G^>(x,y) - G^<(x,y) \right] 
\nonumber \\
&&\quad \times \Bigl\{ 
\ddot\phi + m^2\phi + {\lambda\over 6} \phi^3 
- {i\lambda\over 2} \phi G^>_\alpha(y,y) 
+ {i\lambda^2\over 4} \phi \int_{t_0}^t dt'\, \phi^2(t')
\int d^3\vec z\, \left[ G^>_\alpha(y,z) G^>_\alpha(y,z) 
- G^<_\alpha(y,z) G^<_\alpha(y,z) \right]
\nonumber \\
&&\qquad
- {\lambda\over 2} \phi^2 \int_{t_0}^t dt'
\int d^3\vec z\, \left[ G^>(y,z) - G^<(y,z) \right] \Bigl\{ 
\ddot\phi(t') + m^2\phi(t') + {\lambda\over 6} \phi^3(t')
- {i\lambda \over 2} \phi(t') G^>_\alpha(z,z) 
+ \cdots \Bigr\}
\nonumber \\
&&\qquad
+ {i\lambda\over 2} G^>_\alpha(y,y) \int_{t_0}^t dt' \int d^3\vec z\, 
\left[ G^>(y,z) - G^<(y,z) \right]
\Bigl\{ \ddot\phi(t') + m^2\phi(t') + {\lambda\over 6} \phi^3(t') 
- {i\lambda\over 2} \phi(t') G^>_\alpha(z,z) + \cdots \Bigr\}
\nonumber \\
&&\qquad
+ {\lambda^2\over 4} \phi(t) \int_{t_0}^t dt' \int d^3\vec z\, 
G^>_\alpha(z,z) \left[ G^>_\alpha(y,z) G^>_\alpha(y,z) 
- G^<_\alpha(y,z) G^<_\alpha(y,z) \right]
\nonumber \\
&&\qquad
+ {\lambda^2\over 6} \int_{t_0}^t dt' \int d^3\vec z\, 
\phi(t')  \left[ G^>_\alpha(y,z) G^>_\alpha(y,z) G^>_\alpha(y,z) 
- G^<_\alpha(y,z) G^<_\alpha(y,z) G^<_\alpha(y,z) \right]
\nonumber \\
&&\qquad
+ \cdots \Bigr\}
\label{secondorder}
\end{eqnarray}
\end{widetext}
where $x=(t_f,\vec x)$, $y=(t,\vec y)$ and $z=(t',\vec z)$.  The Wightman functions, $G^{>,<}_\alpha(x,y)$, are defined by 
\begin{eqnarray}
G^>_\alpha(y,z) &=& i\int {d^3\vec k\over 2\omega_k(2\pi)^3}\, 
e^{i\vec k\cdot (\vec y-\vec z)}
\nonumber \\
&&\quad\times
\left[ e^{-i\omega_k(t-t')} 
+ e^{\alpha_k} e^{i\omega_k(2t_0-t-t')} \right] 
\nonumber \\
G^<_\alpha(y,z) &=& i\int {d^3\vec k\over 2\omega_k(2\pi)^3}\, 
e^{i\vec k\cdot (\vec y-\vec z)}
\nonumber \\
&&\quad\times
\left[ e^{i\omega_k(t-t')} 
+ e^{\alpha_k} e^{i\omega_k(2t_0-t-t')} \right] . 
\qquad
\label{aWights} 
\end{eqnarray}
Those without the $\alpha$ subscript are the vacuum mode Wightman functions (for $e^{\alpha_k}=0$) given in Eq.~(\ref{awight}) of the Appendix.  The diagrams for the leading order corrections to the mass and the coupling are those associated with the first line within the braces and are shown in Fig.~\ref{leadloops}.  The diagrams for the remaining terms in Eq.~(\ref{secondorder}) are shown in Fig.~\ref{moreloops}.
In these figures, a solid line represents the fluctuating part of the field, $\psi$, while a dashed line indicates the zero mode, $\phi$.  The shaded blob represents an insertion of the operator, 
\begin{equation}
\ddot\phi + m^2\phi + {\textstyle{1\over 6}} \lambda \phi^3 , 
\label{blob}
\end{equation}
which acts on the outgoing zero mode, as shown in Fig.~\ref{opphi}.  

To calculate the mass renormalization to first order in $\lambda$ and the coupling renormalization to second order, it is sufficient to set the first line in the braces to zero.  The second line corresponds to a self-energy correction to one of the external legs, as can be seen in the figure, while the third, forth and fifth lines are order $\lambda^2$ self-energy corrections.  To the order we shall calculate, we only need to consider mass and coupling renormalization, 
\begin{equation}
Z_0 = 1 + {\cal O}(\lambda), 
\quad 
Z_1 = 1 + {\cal O}(\lambda), 
\quad 
Z_3 = 1 + {\cal O}(\lambda^2), 
\label{Zords}
\end{equation}
since the leading correction to wave function renormalization is order $\lambda^2$ and is from the two-loop contribution given by the fifth line within the braces of Eq.~(\ref{secondorder}).

Setting the integral for the first line of the integrand in Eq.~(\ref{secondorder}) to zero, we find the effect of the interactions on the equation of motion for the zero mode to the specified order.  Substituting in the form of the boundary propagator, Fig.~\ref{leadloops} then gives
\begin{eqnarray}
0 &\!=\!& - \int_{t_0}^{t_f} dt\, {\sin[m(t_f-t)]\over m}
\nonumber \\
&&\times\biggl\{
\ddot\phi(t) 
+ \phi(t) \biggl[ m^2 + {\lambda\over 2} 
\int {d^3\vec k\over (2\pi)^3}\, {1\over 2\omega_k} \biggr]
+ {\lambda\over 6} \phi^3(t) 
\nonumber \\
&&
- {\lambda^2\over 8} \phi(t) \int_{t_0}^t dt'\, \phi^2(t')
\int_0^\infty {d^3\vec k\over (2\pi)^3} 
{\sin(2\omega_k(t-t'))\over\omega_k^2} 
\nonumber \\
&&
+ {\lambda\over 2} \phi(t) 
\int {d^3\vec k\over (2\pi)^3}\, {1\over 2\omega_k} e^{\alpha_k} e^{-2i\omega_k(t-t_0)} 
\nonumber \\
&&
- {i\lambda^2\over 8} \phi(t) \int_{t_0}^t dt'\, \phi^2(t')
\int_0^\infty {d^3\vec k\over (2\pi)^3} 
{e^{\alpha_k}\over\omega_k^2} 
\nonumber \\
&&\qquad\qquad\times 
\left[ e^{-2i\omega_k(t-t_0)} - e^{-2i\omega_k(t'-t_0)} \right] 
\nonumber \\
&& 
+ \cdots \biggr\}. 
\label{zmodeEOM2}
\end{eqnarray}
The first two lines of the integrand are purely bulk effects and are present even for the Poincar\' e-invariant vacuum state.  The last two terms depend explicitly on the boundary conditions, indicated by the factors of $e^{\alpha_k}$.  We treat the renormalization of each set of terms separately, discussing the more familiar bulk renormalization first.

\subsection{Bulk renormalization}

The bulk divergences occur at an arbitrary time in the evolution, so for the purpose of isolating and cancelling these divergences, the modified equation of motion for the zero mode, the integrand of Eq.~(\ref{zmodeEOM2}), is sufficient.  However, unlike the standard nonequilibrium calculation, we must be more careful since the theory contains initial-time divergences.  In particular, the zero mode equation of motion contains spurious divergences that vanish upon integrating its product with the external propagator leg.  Such divergences do not lead to $\mu^\epsilon/\epsilon$ poles in the matrix element---when dimensionally regulating the theory, for example---and thus do not affect the running of the initial conditions.  

In the renormalization of the bulk and boundary effects, we encounter many integrals of a similar general form,
\begin{equation}
I(0,\alpha) = \int {d^3\vec k\over (2\pi)^3}\, {1\over\omega_k^\alpha}
= \int {d^3\vec k\over (2\pi)^3}\, {1\over [\vec k^2 + m^2]^{\alpha/2}} .
\label{drgen0}
\end{equation}
When $\alpha\le 3$, these loop integrals are potentially divergent and we can use dimensional regularization to extract the divergent and finite parts.  Note that since we are only integrating over the spatial momenta, the integral is already Euclidean and no Wick rotation is needed, 
\begin{eqnarray}
I(\epsilon,\alpha) 
&=& \int {d^{3-2\epsilon}\vec k\over (2\pi)^{3-2\epsilon}}\, 
{\mu^{2\epsilon}\over\omega_k^\alpha}
\nonumber \\
&=& {\sqrt{\pi}\over 8\pi^2} {\Gamma(\epsilon - {3-\alpha\over 2})\over
\Gamma({\alpha\over 2})} \left[ {4\pi\mu^2\over m^2} \right]^\epsilon m^{3-\alpha} . 
\label{drgen}
\end{eqnarray}
Here we have included a mass scale $\mu$ to keep the coupling dimensionless, $\lambda \to \mu^{2\epsilon}\lambda$.

The leading tadpole correction to the mass is from the term,
\begin{equation}
{\lambda\over 4} \int {d^3\vec k\over (2\pi)^3}\, {1\over\omega_k}
= - {\lambda m^2\over 32\pi^2} \left[
{1\over\epsilon} + 1 - \gamma + \ln{4\pi\mu^2\over m^2} 
\right] , 
\label{massdiv}
\end{equation}
while the leading correction to the coupling is through 
\begin{equation}
- {\lambda^2\over 8} \phi(t) \int_{t_0}^t dt'\, \phi^2(t')
\int {d^3\vec k\over (2\pi)^3}\, {\sin(2\omega_k(t-t'))\over\omega_k^2} . 
\label{couplecor}
\end{equation}
To extract the divergent piece, integrate by parts with respect to $t'$, 
\begin{eqnarray}
&&\!\!\!\!\!\!\!\!\!\!\!\!\!\!\!\!\!\!\!\!\!\!\!\!
- {\lambda^2\over 8} \phi(t) \int_{t_0}^t dt'\, \phi^2(t')
\int {d^3\vec k\over (2\pi)^3}\, {\sin(2\omega_k(t-t'))\over\omega_k^2} 
\nonumber \\
&=& 
- {\lambda^2\over 16} \phi^3(t) K(0) 
+ {\lambda^2\over 16} \phi(t) \phi^2(t_0) K(t-t_0) 
\nonumber \\
&&
+ {\lambda^2\over 8} \phi(t) \int_{t_0}^t dt'\, \phi(t') \dot\phi(t') K(t-t'),
\label{couplibp}
\end{eqnarray}
where we have defined 
\begin{equation}
K(t-t') 
= \int {d^3\vec k\over (2\pi)^3}\, {\cos(2\omega_k(t-t'))\over\omega_k^3} . 
\label{kerneldef}
\end{equation}
This kernel function diverges logarithmically when its argument vanishes; since the divergence is only logarithmic, the third term on the right side of Eq.~(\ref{couplibp}) is finite upon integration but the first two terms are divergent---the first for an arbitrary $t$ and the second only at $t=t_0$.  Since the first term is proportional to $\phi^3(t)$, it corresponds to a divergent correction to the coupling and is responsible for the familiar running of $\lambda$.  Applying the dimensional regularization result for $\alpha=3$ in Eq.~(\ref{drgen}), $K(0)$ is 
\begin{equation}
K(0) = \int {d^3\vec k\over (2\pi)^3}\, {1\over\omega_k^3} 
= {1\over 4\pi^2} \left[ {1\over\epsilon} - \gamma + \ln {4\pi\mu^2\over m^2} \right] . 
\label{K0div}
\end{equation}
Collecting the results of performing the loop integrals for the standard bulk terms for the finite time-evolved matrix element, we obtain 
\begin{widetext}
\begin{eqnarray}
0 &=& \langle \alpha_k(t_f) | \psi^+(x) | \alpha_k(t_f) \rangle
\nonumber \\
&=&
- \int_{t_0}^{t_f} dt\, {\sin[m(t_f-t)]\over m}
\nonumber \\
&&\quad\times\biggl\{
\ddot\phi(t) 
+ m^2 \phi(t) \left[ 1 - {\lambda\over 16\pi^2} \ln{\mu\over m} 
- {\lambda\over 32\pi^2} \left( {1\over\epsilon} + 1 - \gamma + \ln 4\pi \right) \right]
+ {\lambda\over 4} \phi(t) \int {d^3\vec k\over (2\pi)^3}\, {e^{\alpha_k}\over\omega_k} e^{-2i\omega_k(t-t_0)} 
\nonumber \\
&&\qquad
+ {1\over 6} \phi^3(t) \left[ \lambda - {3\lambda^2\over 16\pi^2} \ln{\mu\over m} 
- {3\lambda^2\over 32\pi^2} \left( {1\over\epsilon} - \gamma + \ln 4\pi \right) \right]
+ {\lambda^2\over 8} \phi(t) \int_{t_0}^t dt'\, \phi(t')\dot\phi(t') K(t-t') 
\nonumber \\
&&\qquad
+ {\lambda^2\over 8} \phi(t) \phi^2(t_0) K(t-t_0) 
- {i\lambda^2\over 8} \phi(t) \int_{t_0}^t dt'\, \phi^2(t') 
\int {d^3\vec k\over (2\pi)^3}\, {e^{\alpha_k}\over\omega_k^2} 
\left[ e^{-2i\omega_k(t-t_0)} - e^{-2i\omega_k(t'-t_0)} \right] 
+ \cdots \biggr\} . 
\qquad
\label{zmodeEOM21}
\end{eqnarray}
\end{widetext}

We now rescale the bare parameters of the theory, as in Eq.~(\ref{Zdefs}), to absorb the $\epsilon\to 0$ divergences.  To order $\lambda$, only $Z_0$ contributes to the mass renormalization, 
\begin{equation}
Z_0 = 1 + {\lambda\over 32\pi^2} 
\left[ {1\over\epsilon} + 1 - \gamma + \ln 4\pi \right] , 
\label{Z0first}
\end{equation}
and only $Z_1$ contributes to the coupling renormalization, 
\begin{equation}
Z_1 = 1 + {3\lambda\over 32\pi^2} 
\left[ {1\over\epsilon} - \gamma + \ln 4\pi \right] , 
\label{Z1first}
\end{equation}
where we have applied the $\overline{\rm MS}$ renormalization scheme.  In terms of the renormalized parameters and neglecting terms of higher order in $\lambda_R$, Eq.~(\ref{zmodeEOM21}) becomes 
\begin{eqnarray}
0 &=& \langle \alpha_k^R(t_f) | \psi^+_R(x) | \alpha_k^R(t_f) \rangle
\nonumber \\
&=& 
- \int_{t_0}^{t_f} dt\, {\sin[m_R(t_f-t)]\over m_R}
\nonumber \\
&&\ \times\biggl\{
\ddot\phi(t) 
+ m_R^2 \phi(t) \left[ 1 - {\lambda_R\over 16\pi^2} \ln{\mu\over m_R} \right]
\nonumber \\
&&\qquad
+ {\lambda_R\over 6} \phi^3(t) \left[ 1 - {3\lambda_R\over 16\pi^2} \ln{\mu\over m_R} \right]
\nonumber \\
&&\qquad
+ {\lambda_R^2\over 8} \phi(t) \int_{t_0}^t dt'\, \phi(t')\dot\phi(t') K(t-t') 
\nonumber \\
&&\qquad
+ {\lambda_R\over 8} \phi(t) \phi^2(t_0) K(t-t_0) 
\nonumber \\
&&\qquad 
+ {\lambda_R\over 4} \phi(t) \int {d^3\vec k\over (2\pi)^3}\, {e^{\alpha_k}\over\omega_k} e^{-2i\omega_k(t-t_0)} 
\nonumber \\
&&\qquad
- {i\lambda^2_R\over 8} \phi(t) \int_{t_0}^t dt'\, \phi^2(t') 
\int {d^3\vec k\over (2\pi)^3}\, {e^{\alpha_k}\over\omega_k^2} 
\nonumber \\
&&\qquad\qquad\qquad\times
\left[ e^{-2i\omega_k(t-t_0)} - e^{-2i\omega_k(t'-t_0)} \right] 
\nonumber \\
&&\qquad
+ \cdots \biggr\} . 
\label{zmodeEOM3}
\end{eqnarray}
The $|\alpha_k^R(t_f)\rangle$ indicates the time-evolved state using the interaction Hamiltonian written in terms of the renormalized parameters.  

The important feature to note from the bulk renormalization so far is that it is, in its short distance behavior, entirely independent of the initial state used.  This behavior is consistent with the principles of effective field theory \cite{eft}.  In choosing the initial state to be other than the standard vacuum, we have not changed the bulk dynamics of the theory and the short distance features of the theory should not know about the state we have chosen once we are sufficiently far from the initial surface.  Defining $\gamma_m(\lambda_R)$ and $\beta(\lambda_R)$ to be 
\begin{equation}
\gamma_m(\lambda_R) = {\mu\over m_R} {d m_R\over d\mu} , 
\qquad
\beta(\lambda_R) = \mu {d\lambda_R\over d\mu} , 
\label{gambetadef}
\end{equation}
we obtain the standard running of the mass and coupling for a $\varphi^4$ theory from Eq.~(\ref{zmodeEOM3}), 
\begin{equation}
\gamma_m(\lambda_R) = {\lambda_R\over 32\pi^2} , 
\qquad
\beta(\lambda_R) = {3\lambda_R^2\over 16\pi^2} , 
\label{gambetaphi4}
\end{equation}
to leading order in $\lambda_R$.

It might appear that we have neglected a possible divergence from the $K(t-t_0)$ term.  Although this term only diverges at the initial boundary, it is independent of the state and apparently produces a new divergence even for the vacuum state.  Since the divergence is only logarithmic, we actually obtain a finite result upon performing the final $dt$ integral.  This term provides a first example of the sort of spurious divergences we must treat carefully when deriving the running of the initial state.  The reason we must neglect these poles in the integrand is that they do not introduce any additional $\mu$ dependence not already present in $\lambda_R$, $m_R$, etc.  For example, the $\mu$-independence of the bare matrix element,
\begin{equation}
\mu {d\over d\mu} 
\langle\alpha_k(t_f) | \psi^+(x) | \alpha_k(t_f) \rangle = 0 ,
\label{baregreens}
\end{equation}
implies that the renormalized matrix element satisfies, 
\begin{eqnarray}
&&\biggl[ \mu {\partial\over\partial\mu} 
+ \beta(\lambda_R) {\partial\over\partial\lambda_R} 
+ \gamma_m(\lambda_R) m_R {\partial\over\partial m_R} 
\nonumber \\
&&\qquad
+ \gamma(\lambda_R)
+ \cdots \biggr]
\langle\alpha^R_k(t_f) | \psi_R^+(x) | \alpha^R_k(t_f) \rangle = 0 ,
\qquad\qquad
\label{rengreens}
\end{eqnarray}
where $\gamma$ is the anomalous dimension and the ellipses refer to the $\mu$-dependence of the boundary conditions.  From this equation, we see that the only important divergences are those which survive {\it all\/} the integrations since only these can produce the $\mu^\epsilon/\epsilon$ terms which affect the renormalization group running.

\section{Boundary renormalization}
\label{bndren}

So far we have considered a completely general spatial dependence for the initial condition, only assuming that this condition is linear in the mode functions.  In this section, we shall isolate the classes of initial conditions that are renormalizable, so it is useful to describe the initial state more systematically.  Let us expand the factor $e^{\alpha_k}$ as a power series in inverse powers of the frequency, $\omega_k$,
\begin{equation}
e^{\alpha_k} = \sum_{n=0}^\infty d_n {m^n\over\omega_k^n} .
\label{ealphser}
\end{equation}
We have expanded in $\omega_k=\sqrt{k^2+m^2}$ rather than $k \equiv |\vec k|$ to avoid the inevitable IR divergences that would occur for an expansion in inverse powers of $k$; in the UV they are essentially the same $\omega_k\sim k$.  For large spatial momenta---the short distance features of the initial state---the $\varpi_k$ which defines the boundary condition in Eq.~(\ref{BDflat}) approaches, 
\begin{equation}
\varpi_k\ {\buildrel k\to\infty\over\longrightarrow}\ 
{1-d_0\over 1+d_0}\, \omega_k 
- {2md_1\over (1+d_0)^2} 
+ {\cal O} \left( {m^2\over\omega_k} \right) , 
\label{asymombar}
\end{equation}
so only the first two terms of the expansion directly distort the infinitesimally short distance features of the state away from the Poincar\' e-invariant state.  Not surprisingly, these two terms produce divergences in the short distance region of loop corrections when evaluated at the initial surface.  Just as for the standard bulk divergences that occur in the perturbative corrections to the mass and coupling, these divergences can be renormalized by adding counterterms, in this case on the initial surface, $t=t_0$.\footnote{Ref.~\cite{dan} presents another approach for addressing initial time divergeces.}  

The moments of the initial conditions with $d_{n\ge 2}$ do have an effect on Green's functions, but since this effect is finite and does not require any renormalization, there is no scale-dependent renormalization associated with these terms.  These states, although they can be quite different from the vacuum at long scales, are only weakly perturbed away from the vacuum in their short distance features.  We have not yet considered any positive powers of $\omega_k$ in Eq.~(\ref{ealphser}) since at short distances such a boundary condition implies that the modes approach the negative energy eigenstates of the free Hamiltonian, $\partial_t U_k(t_0) \approx + i\omega_k U_k(t_0)$.  In this case, the source and image terms effectively reverse their roles in the modes which would be inconsistent with the normalization of the propagator.  Therefore, at infinitesimally short distances, $k\to 0$, positive powers of $\omega_k$ in Eq.~(\ref{ealphser}) become pathological; but as long as we restrict to finite $k<M$, these terms can be treated systematically in $k/M$---such effects form the analogues of the nonrenormalizable operators of a bulk effective field theory.  We shall discuss this class of boundary conditions separately later.

A distinct advantage in stating the boundary conditions in terms of the mode functions, as in Eq.~(\ref{BDflat}) and Eq.~(\ref{ealphser}), is that this approach makes much clearer the relation between the boundary modes and the energy eigenmodes.  Moreover, we also thus avoid the need to specify the detailed form of the boundary Lagrangian that would impose such a condition.  Of course, once we wish to address the boundary renormalization, it becomes convenient to specify a part of the boundary action explicitly---in particular, the counterterms---but we do not want in the process to abandon our mode by mode description in the renormalized theory.  To fix the counterterms in terms of the parameters describing the initial condition---the $d_n$'s---our approach will be to apply a mass-independent renormalization scheme such as $\overline{\rm MS}$ to set the divergent part of the counterterms, as well as the usual set of scale-independent parts that appear in dimensional regularization.  There remains a finite, renormalization scale dependent part of the counterterms, but this part is determined by Callan-Symanzik equation, which also applies to the scale dependence localized on the initial surface.

The divergences on the boundary can be regarded from two vantages.  Their origin lies in looking at the arbitrarily fine details of the state while simultaneously approaching arbitrarily close to the initial time hypersurface.  Therefore, the contribution to the one-point function from boundary effects, contained in the terms 
\begin{eqnarray}
&&\!\!\!\!\!\!\!\!\!\!\!
- \int_{t_0}^{t_f} dt\, {\sin[m(t_f-t)]\over m}
\nonumber \\
&&\times
\biggl\{
{\lambda_R\over 4} \phi(t) \int {d^3\vec k\over (2\pi)^3}\, {e^{\alpha_k}\over\omega_k} e^{-2i\omega_k(t-t_0)} 
\nonumber \\
&&\quad
- {i\lambda_R^2\over 8} \phi(t) \int_{t_0}^t dt'\, \phi^2(t') 
\int {d^3\vec k\over (2\pi)^3}\, {e^{\alpha_k}\over\omega_k^2} 
\nonumber \\
&&\qquad\qquad\qquad\times
\left[ e^{-2i\omega_k(t-t_0)} - e^{-2i\omega_k(t'-t_0)} \right] 
\biggr\} , 
\qquad
\label{zmodeEOMbd}
\end{eqnarray}
is finite as long as {\it either\/} we regulate the momentum integral {\it or\/} we never evaluate the integrand at $t=t_0$.  Initially, it will be useful to consider the latter case since we can thereby establish that all of the new divergent behavior does indeed occur only at the boundary.  Moreover, this regularization allows for a simple determination of the necessary boundary counterterms.  After completing this analysis, we shall then apply the former approach, starting again from Eq.~(\ref{zmodeEOMbd}), by dimensionally regulating the spatial momentum integrals for the terms we have found to contain divergences.

\subsection{Isolating the boundary divergences}

To extract the divergences on the boundary, we proceed in two steps, first determining the $(t-t_0)$-poles in the integrand of Eq.~(\ref{zmodeEOMbd}) before showing that the logarithmic divergences are integrable and thus do not actually produce divergences in the one-point function.  The leading new self-energy correction is from the first term in the integrand of Eq.~(\ref{zmodeEOMbd}), 
\begin{eqnarray}
&&\!\!\!\!\!\!\!\!\!\!\!\!\!\!\!\!\!\!\!\!\!\!\!
{\lambda\over 4} \int {d^3\vec k\over (2\pi)^3}\, {e^{\alpha_k}\over\omega_k} e^{-2i\omega_k(t-t_0)} 
\nonumber \\
&=&
{\lambda\over 8\pi^2} 
\int {k^2\, dk\over\omega_k}\, e^{\alpha_k} e^{-2i\omega_k(t-t_0)} , 
\label{massloopi}
\end{eqnarray}
where we have kept the time arbitrary and integrated over the angular dimensions.  The only parts of this integral that diverge are those associated with the coefficients $d_0$, $d_1$ and $d_2$,
\begin{eqnarray}
&&\!\!\!\!\!\!\!\!\!\!\!\!\!\!
\int_0^\infty {k^2\, dk\over\omega_k}\, 
\left[ d_0 + d_1 {m\over\omega_k} + d_2 {m^2\over\omega_k^2} \right] e^{-2i\omega_k(t-t_0)} 
\nonumber \\
&=& 
{id_0\over 2} {d\over dt} \left( 
\int_0^\infty dk\, e^{-2i\omega_k(t-t_0)} 
\right)
\nonumber \\
&&
+ d_1 m \int_0^\infty dk\, e^{-2i\omega_k(t-t_0)} 
\nonumber \\
&&
+ (d_2-d_0) m^2 \int_0^\infty {dk\over\omega_k}\, e^{-2i\omega_k(t-t_0)} 
\nonumber \\
&&
+ {\rm finite} . 
\label{bnddivL}
\end{eqnarray}
Each of these integrals results in an expression of Hankel functions, $H_\nu^{(2)}$, which only diverge as $t\to t_0$, quadratically,  
\begin{eqnarray}
&&\!\!\!\!\!\!\!\!\!\!\!\!\!\!\!\!\!\!\!
\int_0^\infty dk\, \omega_k e^{-2i\omega_k(t-t_0)} 
\nonumber \\
&=&
{i\over 2} {d\over dt} \left[ 
\int_{-\infty}^\infty d\omega\, 
{\omega\, \Theta(\omega-2m)\over 2\sqrt{\omega^2-4m^2}}
e^{-i\omega(t-t_0)} \right] 
\nonumber \\
&=& {i\pi\over 2} m^2 \biggl[ H^{(2)}_2[2m(t-t_0)] 
- {H^{(2)}_1[2m(t-t_0)]\over 2m(t-t_0)} \biggr]
\nonumber \\
&=& - {1\over 4} {1\over (t-t_0)^2} + \cdots , 
\label{bnddiv2}
\end{eqnarray}
linearly, 
\begin{eqnarray}
\int_0^\infty dk\, e^{-2i\omega_k(t-t_0)} 
&=&
\int_{-\infty}^\infty d\omega\, 
{\omega\, \Theta(\omega-2m)\over 2\sqrt{\omega^2-4m^2}} 
e^{-i\omega(t-t_0)} 
\nonumber \\
&=& - {\pi\over 2} m H^{(2)}_1[2m(t-t_0)] 
\nonumber \\
&=& - {i\over 2} {1\over t-t_0} + \cdots , 
\label{bnddiv1}
\end{eqnarray}
or logarithmically , 
\begin{eqnarray}
\int_0^\infty {dk\over\omega_k}\, e^{-2i\omega_k(t-t_0)} 
&=&
\int_{-\infty}^\infty d\omega\, 
{\Theta(\omega-2m)\over\sqrt{\omega^2-4m^2}} 
e^{-i\omega(t-t_0)} 
\nonumber \\
&=& - {i\pi\over 2} H^{(2)}_0[2m(t-t_0)] 
\nonumber \\
&=& - \ln[m(t-t_0)] + \cdots . 
\label{bnddiv0}
\end{eqnarray}

The second integrand in Eq.~(\ref{zmodeEOMbd}), corresponding to a new correction to the coupling, does not in fact diverge.  The extra power of $\omega_k$ in the denominator of the momentum integral leads only to simple poles and logarithms, 
\begin{eqnarray}
&&\!\!\!\!\!\!\!\!\!\!\!\!\!\!\!\!\!
- {\lambda_R^2\over 32\pi^2} \phi(t) \int_{t_0}^t dt'\, \phi^2(t') 
\nonumber \\
&&\times\biggl\{ 
{d_0\over t-t_0} - {d_0\over t'-t_0} 
- 2i d_1 m \ln {t-t_0\over t'-t_0} 
+ \cdots \biggr\} \quad
\label{coupdivr1}
\end{eqnarray}
where we have used Eqs.~(\ref{bnddiv1}--\ref{bnddiv0}).  This term still contains an additional time integral.  To isolate the behavior as $t\to t_0$, we assume that the zero mode is smooth and finite at $t_0$ so that we can perform a Taylor expansion near the boundary\footnote{Products of logarithms and powers of $(t-t_0)$ also only have a finite effect.}
\begin{equation}
\phi(t) = \sum_{n=0}^\infty {1\over n!} 
\left[ {d^n\phi\over dt^n} \right]_{t=t_0} (t-t_0)^n . 
\label{taylorphi}
\end{equation}
With this assumption, we find that at $t=t_0$, the contribution from Eq.~(\ref{coupdivr1}) is completely finite, 
\begin{eqnarray}
\lim_{t\to t_0}\biggl\{
{i\lambda_R^2\over 8} \phi(t) \int_{t_0}^t dt'\, \phi^2(t') 
\int {d^3\vec k\over (2\pi)^3}\, 
&&
\nonumber \\
{e^{\alpha_k}\over\omega_k^2} 
\left[ e^{-2i\omega_k(t-t_0)} - e^{-2i\omega_k(t'-t_0)} \right] 
\biggr\} 
&\to& {\lambda_R^2d_0 \over 32\pi^2} \phi^3(t_0) . 
\nonumber \\
&&
\label{coupdivr2}
\end{eqnarray}
This result is consistent with our expectation from a dimensional analysis of the operators which could appear in the boundary action.  The $\phi^3(t_0)$ factor is associated with the operator $\phi^3\psi^\pm$ on the three-dimensional boundary theory.  The fields inherit their dimensions from the full $3+1$ dimensional bulk theory, so such an operator is always irrelevant on the boundary.  In a theory with a cubic interaction, the analogous operator $\phi^2\psi^\pm$ would be marginal so additional boundary renormalization would be expected in such a theory.

Thus we have isolated the possible sources of divergences at the boundary which could require renormalization---they appear in either the new one-loop self-energy correction already mentioned,
\begin{equation}
{\lambda_R\over 4} \phi(t) \int {d^3\vec k\over (2\pi)^3}\, {e^{\alpha_k}\over\omega_k} e^{-2i\omega_k(t-t_0)} , 
\label{selfloopenc}
\end{equation}
or in the term which arose from an integration by parts, 
\begin{equation}
{\lambda_R^2\over 16} \phi(t) \phi^2(t_0) K(t-t_0) . 
\label{ibpdebris}
\end{equation}
Applying the Taylor expansion of the field once again, the pole structure derived in Eqs.~(\ref{bnddiv2}--\ref{bnddiv0}) implies that as the integrand approaches $t\to t_0$, 
\begin{eqnarray}
&&\!\!\!\!\!\!\!\!\!\!\!\!\!\!\!\!\!\!\!\!\!\!
{\lambda_R\over 4} \phi(t) \int {d^3\vec k\over (2\pi)^3}\, {e^{\alpha_k}\over\omega_k} e^{-2i\omega_k(t-t_0)} 
\nonumber \\
&\to& 
- {\lambda_R\over 32\pi^2} {1\over t-t_0} d_0\, \dot\phi(t_0)
\nonumber \\
&&
- {\lambda_R\over 32\pi^2} \phi(t_0)
\biggl[ {d_0\over (t-t_0)^2}
+ {2imd_1\over t-t_0} 
\biggr]
\nonumber \\
&& 
- {\lambda_R m_R^2\over 8\pi^2} (d_2-d_0) \phi(t_0) \ln(t-t_0) 
+ \cdots  \quad
\label{selfloopcts}
\end{eqnarray}
for Eq.~(\ref{selfloopenc}) and 
\begin{eqnarray}
&&\!\!\!\!\!\!\!\!\!\!\!\!\!\!\!\!\!\!\!\!\!\!\!
{\lambda_R^2\over 16} \phi(t) \phi^2(t_0) K(t-t_0) 
\nonumber \\
&\to& - {\lambda_R^2\over 32\pi^2} \phi^3(t_0) \ln(t-t_0) + \cdots 
\label{ibpdebriscts}
\end{eqnarray}
for Eq.~(\ref{ibpdebris}).  

Each of these logarithmically diverging terms would be troublesome were their divergences to survive the final time-integration and produce a divergence in the one-point function.  In the latter, Eq.~(\ref{ibpdebriscts}), the $\phi^3(t_0)$ factor indicates that it would require an irrelevant counterterm which would violate our expectation based on a na\"\i ve dimensional analysis of the boundary operators.  Moreover, this term occurs independently of the initial conditions and would imply a boundary renormalization even for the standard vacuum.  In the former logarithmic term of Eq.~(\ref{selfloopcts}), the $d_2$ term has a vanishing effect on the short distance properties of the initial state, compared with the vacuum, so it too should not require any renormalization.  It is therefore important to show that a general contribution of the form, 
\begin{equation}
f(t)\ln(t-t_0) , 
\label{logeg}
\end{equation}
where $f(t)$ and its derivative are smooth and finite at $t=t_0$, has only a finite effect on the one-point function,
\begin{eqnarray}
&&\!\!\!\!\!\!\!\!\!\!\!\!\!\!\!\!
\langle\alpha_k(t)| \psi^+(x) | \alpha_k\rangle 
\nonumber \\
&=&
- \int_{t_0}^{t_f} dt\, {\sin[m(t_f-t)]\over m} 
\Bigl\{ f(t)\ln(t-t_0) + \cdots \Bigr\} . 
\nonumber \\
&&
\label{logtolrp}
\end{eqnarray}
Expanding the argument of the external leg as $t_f-t=(t_f-t_0)-(t-t_0)$ and integrating once by parts yields,
\begin{eqnarray}
&&\!\!\!\!\!\!\!\!\!\!
- \int_{t_0}^{t_f} dt\, {\sin[m(t_f-t)]\over m} f(t)\ln(t-t_0) 
\nonumber \\
&\!=\!& 
{\cos[m(t_f-t_0)]\over m} \int_{t_0}^{t_f} dt\, 
f(t)\ln(t-t_0) \sin[m(t-t_0)] 
\nonumber \\
&&
- f(t_f)\ln(t_f-t_0) {\sin^2[m(t_f-t_0)]\over m^2} 
\nonumber \\
&&
+ {\sin[m(t_f-t_0)]\over m^2} \int_{t_0}^{t_f} dt\, 
\dot f(t)\ln(t-t_0) \sin[m(t-t_0)] 
\nonumber \\
&&
+ {\sin[m(t_f-t_0)]\over m^2} \int_{t_0}^{t_f} dt\, 
f(t) {\sin[m(t-t_0)]\over t-t_0} . 
\nonumber \\
&& 
\label{logtolrpint}
\end{eqnarray}
Each of these terms is finite when $f(t)$ and its derivative are sufficiently well-behaved.  What this analysis shows is that while terms such as that in Eq.~(\ref{ibpdebriscts}) and in the last line of Eq.~(\ref{selfloopcts}) contain logarithmic divergences, these effects only have a finite effect on the integrated equation of motion.  Therefore, the logarithm terms in Eq.~(\ref{selfloopcts}) and Eq.~(\ref{ibpdebriscts}) do not require infinite counterterms and consequently they have no dependence on the renormalization scale $\mu$.

Only two of the moments which define the initial conditions require regularization, $d_0$ and $d_1$.  Reading off the dependence on the zero mode $\phi(t)$ in the terms that diverge on the boundary in Eq.~(\ref{selfloopcts}), the counterterms needed are 
\begin{equation}
S_{t=t_0} = \int_{t=t_0} d^3\vec y\, {\textstyle{1\over 2}} \left\{ 
z_0 \dot\phi\psi^\pm 
+ z_0 \phi\partial_t\psi^\pm
- z_1 m \phi\psi^\pm 
\right\} . 
\label{surfactfA}
\end{equation}
which can be equivalently written as a bulk contribution, 
\begin{eqnarray}
S_{t=t_0} 
&=& - \int_{t_0}^\infty dt \int d^3\vec y\, \Bigl\{ 
{\textstyle{1\over 2}} z_0 \dot\delta(t-t_0)\, \phi\psi^\pm
\nonumber \\
&&\qquad\qquad\qquad\quad 
+ {\textstyle{1\over 2}} z_1 m \delta(t-t_0)\, \phi\psi^\pm 
\Bigr\} . \qquad\quad
\label{surfactfB}
\end{eqnarray}

\subsection{Boundary dimensional regularization}

Having shown explicitly that the only new divergences are localized on the initial boundary, we reevaluate these terms at $t=t_0$ but instead we use dimensional regularization to extract the divergent pieces since this regularization method allows us to apply the familiar $\overline{\rm MS}$ scheme for cancelling the poles as the number of spatial dimension approaches three.  Because of the extra overall time integration, the dimensional regularization of the boundary divergences is somewhat more subtle than the bulk regularization.  

We begin by integrating by parts until the new self-energy correction is no more than logarithmically divergent, 
\begin{equation}
\int {d^3\vec k\over (2\pi)^3}\, {e^{\alpha_k}\over\omega_k} e^{-2i\omega_k(t-t_0)} 
= - {1\over 4} {d^2\over dt^2} K_\alpha(t-t_0), 
\label{ddKalpha}
\end{equation}
where we have introduced a new kernel function, 
\begin{equation}
K_\alpha(t-t_0) = 
\int {d^3\vec k\over (2\pi)^3}\, {e^{\alpha_k}\over\omega_k^3} e^{-2i\omega_k(t-t_0)} . 
\label{Kalphadef}
\end{equation}
The contribution of this term to the expectation value of $\psi^+(x)$ is then
\begin{equation}
\int_{t_0}^{t_f} dt\, {\sin[m(t_f-t)]\over m} 
\left\{ {\lambda_R\over 16} \phi(t) \ddot K_\alpha(t-t_0) \right\} . 
\label{abbr}
\end{equation}
To simplify the notation, we shall abbreviate the product of the zero mode and the external propagator leg by
\begin{eqnarray}
{\cal G}(t) &\equiv& \int d^3\vec y\, \left[ G^>(x,y) - G^<(x,y) \right] \phi(t)
\nonumber \\
&=& {\sin[m(t_f-t)]\over m} \phi(t) . 
\label{gdef}
\end{eqnarray}
Integrating Eq.~(\ref{abbr}) by parts twice yields, 
\begin{eqnarray}
&&\!\!\!\!\!\!\!\!\!\!\!\!\!
\int_{t_0}^{t_f} dt\, {\cal G}(t) \ddot K_\alpha(t-t_0) 
\nonumber \\
&=& 
- {\cal G}(t_0) \dot K_\alpha(0) + \dot{\cal G}(t_0) K_\alpha(0) 
+ \phi(t_f) K_\alpha(t_f-t_0) 
\nonumber \\
&&
+ \int_{t_0}^{t_f} dt\, \ddot{\cal G}(t) K_\alpha(t-t_0) . 
\label{abbribp}
\end{eqnarray}
The final integrand contains an integrable logarithmic singularity as has already been noted.  

The two surface terms with $K(0)$ and $\dot K(0)$ are divergent and we regulate them using dimensional regularization, as in Eq.~(\ref{drgen}), 
\begin{equation} 
K_\alpha(0) = {d_0\over 4\pi^2} \left[ {1\over\epsilon} - \gamma + \ln {4\pi\mu^2\over m^2} \right] + {\rm finite} 
\label{Kalph0}
\end{equation}
and 
\begin{equation}
\dot K_\alpha(0) = 
{imd_0\over 2\pi} 
- {imd_1\over 2\pi^2} \left[ 
{1\over\epsilon} - \gamma + \ln{4\pi\mu^2\over m^2} \right] 
+ {\rm finite} . 
\label{dKalph0}
\end{equation}
Thus the new boundary divergences correspond to the following ${1\over\epsilon}$ poles, 
\begin{eqnarray}
&&\!\!\!\!\!\!\!\!\!\!\!\!\!
\int_{t_0}^{t_f} dt\, {\lambda_R\over 16} {\cal G}(t) \ddot K_\alpha(t-t_0) 
\nonumber \\
&=& 
\left[ \dot {\cal G}(t_0) {\lambda_R d_0\over 64\pi^2} 
+ {\cal G}(t_0) {im\lambda_R d_1\over 32\pi^2} 
\right] 
\left[ 
{1\over\epsilon} - \gamma + \ln{4\pi\mu^2\over m^2} \right]
\nonumber \\
&&
+ {\rm finite}
\label{abbribpeps}
\end{eqnarray}

To cancel these poles, we add the surface counterterms to the action using Eq.~(\ref{surfactfB}).  The contribution of this action to the interaction Hamiltonian is then
\begin{eqnarray}
&&\!\!\!\!\!\!\!\!\!\!
H_I^{\rm surface}(\phi,\psi^\pm) 
\nonumber \\
&=& - \int d^3\vec y\, \left[ 
{\textstyle{1\over 2}} \dot\delta(t-t_0) z_0\, \phi \psi^\pm 
+ {\textstyle{1\over 2}} \delta(t-t_0) z_1 m_R\, \phi \psi^\pm 
\right] . 
\nonumber \\
&&
\label{HIsurf}
\end{eqnarray}
To leading order in these counterterms, the correction to the equation of motion for the zero mode is 
\begin{eqnarray}
&&\!\!\!\!\!\!\!\!\!\!\!\!\!\!\!\!\!\!\!\!\!\!\!\!\!\!\!
- \int_{t_0}^{t_f} dt\, {\cal G}(t) 
\left[ {\textstyle{1\over 2}} z_0 \dot\delta(t-t_0) 
+ {\textstyle{1\over 2}} z_1 m_R \delta(t-t_0) \right] 
\nonumber \\
&=& 
{\textstyle{1\over 2}} z_0 \dot {\cal G}(t_0) 
- {\textstyle{1\over 2}} z_1 m_R {\cal G}(t_0) . 
\label{abbribpcts}
\end{eqnarray}
Adding together these terms and Eq.~(\ref{abbribpeps}) gives the following surface contribution to the expectation value of the fluctuation, 
\begin{eqnarray}
&=& 
- \dot {\cal G}(t_0) \left\{ {1\over 2} z_0 + {\lambda_R d_0\over 64\pi^2} 
\left[ 
{1\over\epsilon} - \gamma + \ln{4\pi\mu^2\over m^2} \right]
\right\} 
\nonumber \\
&& 
+ {\cal G}(t_0) m_R \left\{ {1\over 2} z_1 - {i\lambda_R d_1\over 32\pi^2} 
\left[ 
{1\over\epsilon} - \gamma + \ln{4\pi\mu^2\over m^2} \right] . 
\right\}
\nonumber \\
&&
\label{cancelsurf}
\end{eqnarray}

The counterterms contain both divergent, $\mu$-independent pieces which are fixed by the $\overline{\rm MS}$ renormalization scheme as well as finite, scale-dependent parts, so we shall separate each of these effects explicitly,
\begin{equation}
z_0 = z_0^\epsilon + \hat z_0(\mu) , 
\qquad
z_1 = z_1^\epsilon + \hat z_1(\mu) . 
\label{fininfin}
\end{equation}
The finite parts are not arbitrary but are determined by the Callan-Symanzik equation as we shall see.  The $\overline{\rm MS}$ scheme fixes 
\begin{eqnarray}
z_0^\epsilon &=& - {\lambda d_0\over 32\pi^2} 
\left[ {1\over\epsilon} - \gamma + \ln 4\pi \right] ,
\nonumber \\
z_1^\epsilon &=& {i\lambda d_1\over 16\pi^2} 
\left[ {1\over\epsilon} - \gamma + \ln 4\pi \right] . 
\label{ctvalues}
\end{eqnarray}
Note that to the order at which we are solving for the surface counterterms, $\lambda = \lambda_R + {\cal O}(\lambda_R^2)$.  

The renormalized one-point function, written in terms of finite parameters therefore becomes
\begin{widetext}
\begin{eqnarray}
0 &=& \langle\alpha_k^R(t_f)| \psi^+_R(x) |\alpha_k^R(t_f)\rangle
\nonumber \\
&=& - \int_{t_0}^{t_f} dt\, {\sin[m_R(t_f-t)]\over m_R}
\biggl\{
\ddot\phi(t) 
+ m_R^2 \phi(t) \left[ 1 - {\lambda_R\over 16\pi^2} \ln{\mu\over m_R} \right] 
+ {\lambda_R\over 6} \phi^3(t) 
\left[ 1 - {3\lambda_R\over 16\pi^2} \ln{\mu\over m_R} \right]
\nonumber \\
&&\qquad\qquad\qquad\qquad\qquad\quad
+ {\lambda_R^2\over 8} \phi(t) \int_{t_0}^t dt'\, \phi(t')\dot\phi(t') K(t-t') 
+ {\lambda_R^2\over 8} \phi(t) \phi^2(t_0) K(t-t_0) 
\biggr\} 
\nonumber \\
&&
+ \dot{\cal G}(t_0) 
\left[ {1\over 2} \hat z_0(\mu) + d_0 {\lambda_R\over 32\pi^2} \ln{\mu\over m_R} \right]
+ {\cal G}(t_0) m_R 
\left[ - {1\over 2} \hat z_1(\mu) + d_1 {i\lambda_R\over 16\pi^2} \ln{\mu\over m_R} \right] 
\nonumber \\
&&
+ {\lambda_R\over 16} \left\{
- {\cal G}(t_0) 
\left[
{im_Rd_0\over 2\pi} + \dot{\hat K_\alpha}(0) 
\right]
+ \dot{\cal G}(t_0) \left[ \hat K_\alpha(0)
\right]
+ \phi(t_f) K_\alpha(t_f-t_0) 
+ \int_{t_0}^{t_f} dt\, \ddot{\cal G}(t) K_\alpha(t-t_0) . 
\right\}
\nonumber \\
&& - {\lambda_R^2\over 16} \int_{t_0}^{t_f} dt\, {\sin[m_R(t_f-t)]\over m_R}
\phi(t) \int_{t_0}^t dt'\, \phi^2(t') 
\left\{ \dot K_\alpha(t-t_0) - \dot K_\alpha(t'-t_0) \right\}
\nonumber \\
&&
+ \cdots . 
\label{matrixmu} 
\end{eqnarray}
\end{widetext}
to order $\lambda_R$ in the self-energy corrections and to $\lambda_R^2$ in the coupling correction.  We have written $\hat K_\alpha$ for the finite kernel function with the $d_0$ and $d_1$ moments removed,
\begin{equation}
\hat K_\alpha(t-t_0) = 
\int {d^3\vec k\over (2\pi)^3}\, \sum_{n=2}^\infty d_n {m^n_R\over\omega_k^{n+3}} e^{-2i\omega_k(t-t_0)} . 
\label{hatKalphadef}
\end{equation}

Here we have been using the one-point function to determine the scale dependence of the parameters of the theory.  It has a further consequence on the zero mode, which in an expanding background affects the expansion.  The usual interpretation is that the zero mode is free to decay into the other degrees of freedom; a signal of this effect is the dissipative $\dot\phi$ term present even in the standard vacuum, $e^{\alpha_k}=0$, which is seen by setting the integrand in Eq.~(\ref{matrixmu}) to zero 
\begin{eqnarray}
0 &=& \ddot\phi 
+ m_R^2 \left[ 1 - {\lambda_R\over 16\pi^2} \ln{\mu\over m_R} \right] 
\nonumber \\
&&
+ {\lambda_R\over 6} \phi^3 
\left[ 1 - {3\lambda_R\over 16\pi^2} \ln{\mu\over m_R} \right]
\nonumber \\
&&
+ {\lambda_R^2\over 8} \phi(t) \phi^2(t_0) K(t-t_0) 
\nonumber \\
&&
+ {\lambda_R^2\over 8} \phi(t) \int_{t_0}^t dt'\, \phi(t')\dot\phi(t') K(t-t') 
\nonumber \\
&&
+ \cdots . 
\label{zmodemod} 
\end{eqnarray}
The presence of many more terms for a general initial condition implies a much more complicated effect on the zero mode equation of motion.  In effect, we no longer have the zero mode decaying in a $\psi$-vacuum, but rather one that decays in a specified background of excited $\psi$-modes.

\subsection{The Callan-Symanzik Equation}

Once we have renormalized both the bulk and the boundary components of the action, we have introduced a dependence on the renormalization scale $\mu$ into a general Green's function.  But since the Green's functions can be alternately expressed in terms of the bare quantities or their rescaled counterparts, they are independent of the renormalization scale---as is familiar from the standard $S$-matrix description of a field theory.  This scale independence is the foundation for the Callan-Symanzik equation and we shall derive here what this equation implies for the running of the boundary effects.

Let us define the renormalized $n$-point connected Green's function for the fluctuations by 
\begin{eqnarray}
&&\!\!\!\!\!\!\!\!\!\!\!\!\!\!\!
G^{(n)}_R(x_1, \ldots , x_n) 
\nonumber \\
&=& \langle \alpha_k^R(t_f) | 
T_\alpha \bigl( \psi^+_R(x_1) \cdots \psi^+_R(x_n) \bigr) | 
\alpha_k^R(t_f) \rangle_{\rm connected} . 
\nonumber \\
&&
\label{connpoint}
\end{eqnarray}
The equivalence of the bare and renormalized forms of this Green's function implies that it satisfies the Callan-Symanzik equation,
\begin{eqnarray}
\mu {d\over d\mu} G^{(n)}_R(x_1, \ldots , x_n) = 0 , 
\label{callan} 
\end{eqnarray}
where the $\mu$ dependence appears in both the bulk parameters as well as in the operators confined to the initial surface which are necessary to renormalize the short distance features of the initial condition.  In particular, the one-point function studied throughout this section obeys
\begin{equation}
\mu {d\over d\mu} G^{(1)}_R(x) = 0 , 
\label{callanone} 
\end{equation}
or 
\begin{eqnarray}
\biggl[ \mu {\partial\over\partial\mu} 
+ \beta(\lambda_R) {\partial\over\partial\lambda_R} 
+ \gamma_m(\lambda_R) m_R {\partial\over\partial m_R} 
\qquad\qquad
&&
\nonumber \\
+ \gamma(\lambda_R) 
+ \mu {d\hat z_0\over d\mu} {\partial\over\partial\hat z_0} 
+ \mu {d\hat z_1\over d\mu} {\partial\over\partial\hat z_1} 
\biggr]
G^{(1)}_R(x) 
&=& 0
\nonumber \\
&&
\label{callanonefull}
\end{eqnarray}
for boundary conditions of the form given in Eq.~(\ref{ealphser}).  Using the form for the one-point function given in Eq.~(\ref{matrixmu})---retaining terms of order $\lambda_R$ which are linear in the zero mode $\phi$ and terms of order $\lambda_R^2$ which are cubic in $\phi$---we find that 
\begin{widetext}
\begin{eqnarray}
0 
&=& - \int_{t_0}^{t_f} dt\, 
\left[ {\partial\over\partial m_R} {\sin[m_R(t_f-t)]\over m_R} \right]
m_R \gamma_m \biggl\{
\ddot\phi(t) + m_R^2 \phi(t) + {\lambda_R\over 6} \phi^3(t) 
\biggr\} 
\nonumber \\
&& - \int_{t_0}^{t_f} dt\, {\sin[m_R(t_f-t)]\over m_R}
\biggl\{
2 m_R^2 \phi(t) 
\left[ \gamma_m(\lambda_R) - {\lambda_R\over 32\pi^2} \right] 
+ {1\over 6} \phi^3(t) 
\left[ \beta(\lambda_R) - {3\lambda_R^2\over 16\pi^2} \right]
\biggr\} 
\nonumber \\
&&
+\ {1\over 2} \dot{\cal G}(t_0) 
\left[ \mu {d\hat z_0\over d\mu} + d_0 {\lambda_R\over 16\pi^2} \right]
- {1\over 2} m_R {\cal G}(t_0) 
\left[ 
\mu {d\hat z_1\over d\mu} - d_1 {i\lambda_R\over 8\pi^2} \right] 
+ \cdots . 
\label{CSonepoint} 
\end{eqnarray}
\end{widetext}
The first line vanishes to this order using the tree-level equation of motion for the zero mode while the second vanishes when the mass and coupling have the standard running.  The third line determines the running of the boundary effects, 
\begin{eqnarray}
\mu {d\hat z_0\over d\mu} 
&=& - {\lambda_R\over 16\pi^2} d_0 + {\cal O}(\lambda_R^2)
\nonumber \\
\mu {d\hat z_1\over d\mu} 
&=& {i\lambda_R\over 8\pi^2} d_1 + {\cal O}(\lambda_R^2) . 
\label{bndrun}
\end{eqnarray}

\section{Nonrenormalizable initial conditions}
\label{nonrenorm}

The boundary conditions we have been considering depart from the standard vacuum most significantly at long distances, but at short distances the state still has a strong resemblance to the vacuum up to an overall rescaling and corrections which diminish as inverse powers of the three momentum.  Not surprisingly, the divergences produced by these terms can be cancelled by, at worst, marginal counterterms confined to the initial boundary.  Thus we discover a correspondence between {\it renormalizable\/} boundary theories and {\it small\/} departures from the vacuum state at short distances.  This correspondence can be extended to relate larger departures from the vacuum state to nonrenormalizable actions on the boundary.

For illustration, let us consider in some detail a boundary whose short distance features progressively differ from the vacuum as 
\begin{equation}
e^{\alpha_k} = {\omega_k\over M} ; 
\label{linbound}
\end{equation}
$M$ is a mass scale and we have absorbed any dimensionless coefficients into the definition of this scale for now.  We shall also assume that $m\ll M$.  This boundary condition is essentially a UV modification of the initial state---at long distances $k\ll M$, $e^{\alpha_k}\to m/M \ll 1$.  At longer and longer scales, its effects become increasing irrelevant to propagating fields in the bulk.  At short distances its behavior becomes pathological.  Notice that as $k\gg M$, the boundary condition approaches that defining the negative energy eigenstate, 
\begin{equation}
\left. \partial_t U_k(t) \right|_{t=t_0} = 
i \omega_k 
\Bigl[ 1 - {2M\over\omega_k} + \cdots \Bigr] U_k(t_0) , 
\label{badbound}
\end{equation}
as mentioned earlier.  Nothing however forbids applying a boundary condition such as Eq.~(\ref{linbound}) as long as we restrict to scales sufficiently below $M$.  For example, if the dynamics that sets the initial condition at $t=t_0$ contains a mass scale $M$, it is natural to expect that the state should inherit features at this scale.  If the preceding dynamics is otherwise local and well behaved at shorter distances, then above this scale the difference between the initial state and the vacuum should again diminish, growing negligible in the far UV.  Thus a boundary condition such as Eq.~(\ref{linbound}) can be very useful for describing some preceding physics with a mass scale $M$; but such a condition is necessarily incomplete since we need to apply another boundary condition for the features of the initial scale that are smaller than $\sim 1/M$.  In this sense, terms in a power series expansion of the boundary condition that scale as some positive power of $\omega_k$ resemble nonrenormalizable operators in an effective field theory---their effects are small in the IR, they cannot be part of a ``full'' UV-consistent theory, but they can be very useful for encoding the effects of that more complete theory if we do not ask too much about the theory and its short distance behavior.

We examine this connection between an IR-irrelevant modification to the initial condition and nonrenormalizable operators on the boundary by re-examining the one-loop correction to the coupling $\lambda_R$ calculated before, but now using the initial condition in Eq.~(\ref{linbound}).  In the integrand of Eq.~(\ref{zmodeEOM2}), the leading cubic term in the zero mode $\phi$ that depends on the initial condition is 
\begin{eqnarray}
&&\!\!\!\!\!\!\!\!\!\!\!\!
-{i\lambda_R^2\over 8} \phi(t) \int_{t_0}^t dt'\, \phi^2(t') 
\nonumber \\
&&\times 
\int {d^3\vec k\over (2\pi)^3} {e^{\alpha_k}\over\omega_k^2} 
\left[ e^{-2i\omega_k(t-t_0)} - e^{-2i\omega_k(t'-t_0)} \right] 
\quad
\label{cubterm}
\end{eqnarray}
where we have not written the external $\psi$-leg.  Applying the boundary condition in Eq.~(\ref{linbound}) and integrating over the spatial momentum as in Eq.~(\ref{bnddiv2}), we see that this loop correction now produces a divergence at $t=t_0$, 
\begin{equation}
= - {i\over 64\pi^2} {\lambda_R^2\over M} 
\left\{ {\phi^3(t_0)\over t-t_0} 
+ 2 \phi^2(t_0) \dot\phi(t_0) + \cdots \right\}
\label{cubtermdiv}
\end{equation}
where the ellipses refer to terms that vanish as $t\to t_0$.  To cancel this divergence requires a counterterm proportional to $\phi^3\psi^\pm$ on the boundary or, written in terms of the full field, 
\begin{equation}
S^{\rm new} = \int_{t=t_0} d^3\vec x\, \left\{ 
{1\over 24} {z_3\over M} \varphi^4 + \cdots \right\} . 
\label{irrelbnd}
\end{equation}
We have included the mass scale $M$ in the boundary action so that $z_3$ remains dimensionless.  Two important properties of this IR-irrelevant boundary condition should be mentioned.  The new divergences to the theory are still confined to the initial surface but now they require counterterms which correspond to irrelevant operators of the boundary theory.  We generically expect the need for further boundary counterterms, such as
\begin{equation}
\varphi^6, \varphi^8, \ldots , 
\label{higherops}
\end{equation}
as well as terms with insertions of arbitrary powers of time derivatives or even powers of spatial derivatives, which appear at higher order in $\lambda_R$.  The complete set of dimension four boundary operators, symmetric under $\varphi\leftrightarrow -\varphi$, is 
\begin{equation}
\varphi^4,\quad \varphi \ddot\varphi,\quad 
\dot\varphi^2,\quad 
\vec\nabla\varphi \cdot \vec\nabla\varphi . 
\label{dimfourops}
\end{equation}

From this example, we now understand how to interpret a more general boundary condition,
\begin{equation}
e^{\alpha_k} = \sum_{n=0}^\infty d_n {m^n\over\omega_k^n} 
+ \sum_{n=1}^\infty c_n {\omega_k^n\over M^n} . 
\label{genbound}
\end{equation}
The first sum contains terms that are important for the long-distance features of the initial state.  They can produce, at least for the first few terms in the series, divergences on the initial boundary which are absent for the vacuum and which are removed by adding relevant or marginal counterterms on the boundary.  We apply an $\overline{\rm MS}$ prescription to cancel the $1/\epsilon$ pole and usual constant finite factors and find the running in $\mu$ of the counterterms through the Callan-Symanzik equation.  The second sum does not appear to make sense, but this apparent pathology only arises if we attempt to use the condition in regime beyond its inherent scale of applicability, which is signaled by the mass scale $M$.  At low scales, set by $\phi(t_0)$ and its derivatives for example, the higher order terms in the second sum are suppressed by additional powers of $M$, and to a given precision, we only need to consider a finite set of these terms.  To obtain a sense of the scales, consider the finite correction for the linear term in $\omega_k$ in Eq.~(\ref{cubtermdiv}) at $t=t_0$ and compare it to the analogous finite correction in Eq.~(\ref{coupdivr2}) for $d_0\not=0$.  The ratio of the former to the latter is 
\begin{equation}
{ - {i\over 32\pi^2} {\lambda_R^2\over M} c_1 \phi^2(t_0) \dot\phi(t_0) 
\over 
{\lambda_R^2\over 32\pi^2} d_0 \phi^3(t_0) }
\sim
{1\over M} {\dot\phi(t_0)\over\phi(t_0)} , 
\label{ratioIRUV}
\end{equation}
so when the derivative of the zero mode is sufficiently small on the boundary, this effect will be suppressed.  Higher order terms in $1/M$ will be consequently even more suppressed in this regime.  In an expanding background, we have the additional scale for the rate of expansion, the Hubble scale $H$, so the natural suppression factor is $H/M$.

We are now ready to describe the prescription for treating the terms in the second sum.  These terms also generically produce divergences which are not present in the vacuum but these too are confined to the initial surface.  The divergences are removed by applying the boundary-$\overline{\rm MS}$ scheme again, this time adding irrelevant counterterms on the boundary, and the counterterms also contain finite parts that run in $\mu$.  The theory remains predictive as long as we only demand a finite accuracy of it---higher order corrections are suppressed by $\Delta/M$, where $\Delta$ can any of the possible scales in the problem, 
\begin{equation}
\Delta \sim \phi(t_0),\ {\dot\phi(t_0)\over\phi(t_0)},\ 
|\vec k|,\ {1\over t-t_0},\ \ldots .
\label{equation}
\end{equation}
The last of these quantities emphasizes that we should not expect the a controlled expansion arbitrarily close to the initial surface, $t\to t_0$, since this also corresponds to a UV limit.  For $\Delta\ll M$, we only require the leading terms of the second sum in Eq.~(\ref{genbound}) in practice.

Since this class of boundary conditions generically requires some boundary renormalization, we can expect that they produce effects that run with the renormalization scale.  This running is determined by appropriately extending the Callan-Symanzik equation in Eq.~(\ref{callanonefull}) so that it fixes the $\mu$ dependence of the finite parts of the coefficients that accompany the boundary counterterms.  In a cosmological setting, this running means that corrections can be enhanced by logarithmic factors---$(H/M)\ln(H/M)$ rather than $H/M$ order corrections.

\subsection{Na\"\i ve power counting}

Some care must be taken when estimating of the degree of the boundary divergence for a general loop contribution.  In particular, a na\"\i ve approach based on the $S$-matrix typically overestimates the degree of divergence.  As an example, consider once again the order $\lambda_R^2$ correction to the coupling in the tadpole given in Eq.~(\ref{cubterm}) and represented by the third graph in Fig.~\ref{leadloops}.  The loop contains two propagators so that it might appear that its boundary divergence for large values of the spatial momentum should scale as
\begin{equation}
d^3\vec k \times {1\over\omega_k^2} \times e^{2\alpha_k} 
\sim k e^{2\alpha_k}
\label{nda}
\end{equation}
where the factors on the left side correspond to the measure of the loop integral, two powers of $\omega_k^{-1}$ for the two propagators and two powers of the boundary factor $e^{\alpha_k}$ for the boundary-dependent parts of these propagators.  Thus for the linearly scaling example above, this graph diverges as $k^3/M^2$, although the final degree of divergence can be reduced by the time integrals as we have seen.  However, in Eq.~(\ref{cubterm}) only one power of $e^{\alpha_k}$ appears.  In the Schwinger-Keldysh approach, the contribution from this graph, again omitting the external leg, is 
\begin{eqnarray}
&& {i\lambda^2\over 4} \phi(t) \int_{t_0}^t dt'\, \phi^2(t')
\nonumber \\
&&\qquad\times
\int d^3\vec z\, \left[ G^>_\alpha(y,z) G^>_\alpha(y,z) 
- G^<_\alpha(y,z) G^<_\alpha(y,z) \right] 
\nonumber \\
&&
\label{sndorder}
\end{eqnarray}
Since the boundary dependent part of the two-point functions $G^>_\alpha(y,z)$ and $G^<_\alpha(y,z)$ is identical, given in Eq.~(\ref{aWights}), the order $\left( e^{\alpha_k} \right)^2$ pieces cancel between the two terms.  Thus the divergence is milder than what Eq.~(\ref{nda}) would have na\"\i vely predicted.

\section{Conclusions}
\label{conclude}

An effective theory description of an initial state provides a powerful formalism for understanding and quantifying the observability of the very short distance features of that state.  Such a framework is especially needed in inflation where the Hubble scale could lie far beyond the scales probed by accelerator experiments, while the states of the fields at the beginning of inflation, which are presumably set by some preceding dynamics, are unknown.  Just as for an ordinary field theory in the bulk space-time, the features of the initial state can be characterized as renormalizable and nonrenormalizable and it is in the latter that the effects of the ``trans-Planckian'' physics are encoded.  The scaling of their observable signature is completely determined by the effective description of the initial state, without the need to appeal to a particular model for the details near the Planck scale.

Even in flat space, choosing an initial state other than the vacuum affects the structure of the propagator so that it includes an extra ``image'' term encoding the propagation of initial state information.  This additional structure is necessary for the consistency with the initial state.  Moreover, for states that differ sufficiently from the vacuum at short distances, this structure is also needed for the renormalizability of the theory and can be seen as a consequence of correctly time-ordering operators to prevent uncontrolled divergences in the bulk theory \cite{lowe,einhorn,taming}.  The theory can nevertheless contain divergences beyond those present for the vacuum.  These divergences arise directly out of the short distance structure of the initial state and they occur exactly at the initial time surface at which the state is defined.  Once the short distance behavior of the initial state has been renormalized at this surface, the theory remains finite for all subsequent times.

Renormalizable initial conditions are characterized as those that still yield the same short distance behavior as the vacuum state up to possible constant rescalings.  The new divergences associated with this class of initial conditions can be renormalized through relevant or marginal---with respect to the boundary theory---counterterms confined to the initial boundary.  Nonrenormalizable initial conditions are those differing significantly from the vacuum condition at short distances.  It might be thought that they are therefore unphysical, but they are predictive in the usual sense of an effective theory.  If these effects become large on scales of order $M$, then as long as we only probe up to a scale $\Delta\ll M$ there is only a finite set of parameters needed to describe the nonrenormalizable features of the state up to a given order in $(\Delta/M)^n$.  Here too we have divergences at the initial surface from summing over more and more of the short distance features of the initial state; such divergences are cancelled by irrelevant, local operators also confined to the initial surface.  The Callan-Symanzik equation determines the renormalization group running associated with these operators, which applies to both the bulk and boundary renormalization.

In principle, the renormalization of an initial state should proceed very similarly in an expanding background.  The reason lies in how the standard vacuum state is defined in a Robertson-Walker universe.  The state is assumed to be invariant under all of the isometries of the background, which with the exception of de Sitter space are always fewer in number those of Minkowski space.  Even in de Sitter space \cite{alpha}, this requirement does not select a unique state and an additional condition must be imposed.  For example, in de Sitter space the Bunch-Davies \cite{bunch} vacuum is that invariant state whose modes match with the Minkowski space vacuum modes at scales much smaller than the horizon, $1/H$.  A similar prescription applies also to a Robertson-Walker universe.

Thus, in terms of its short distance features, how a state differs from the vacuum in a curved space-time is essentially the same as in flat space.  In the expanding case, the curvature scale, $H$, defines the relevant scale for our effective theory. Choosing a spatially flat form for the Robertson-Walker metric, the short distance features of the modes can be written as
\begin{equation}
U_k^\alpha(t) \to e^{-ikt} + e^{\alpha_k} e^{i\delta_k} e^{ikt} 
\qquad
{\rm as}
\quad
k=|\vec k|\gg H 
\label{frwflatlim}
\end{equation}
up to possible relative phases, $e^{i\delta_k}$.  The vacuum state is defined by choosing only the ``positive energy'' modes, $e^{\alpha_k}=0$; but for a general initial state we should include the conjugate term as well.  As $k\gg H$, we can write
\begin{equation}
e^{\alpha_k} = \hbox{IR-relevant moments} 
+ \sum_{n=1}^\infty c_n {k^n\over M^n} , 
\label{genUV}
\end{equation}
just as before.  For the effective description of the state to be predictive, the scale at which the difference between the initial state and the vacuum, $M$, becomes important should be sufficiently smaller than $H$ so that the contributions of the higher order moments will be suppressed by $(H/M)^n$.  We shall explore the renormalization of nonvacuum initial conditions more extensively in \cite{initfrw}.

The effective theory framework provides the natural setting in which to address many of the mysterious aspects of inflation \cite{branpascos}.  The basis of the effective theory idea is to retain only the relevant physics for the scale being studied.  As we have seen, the ultraviolet physics is still present but if we have set up our effective theory correctly, these UV effects are both small and controlled at this scale.  The idea also applies to extreme infrared effects. In particular, we should not need to know details on  superhorizon scales to make predictions within our own causal horizon.  From this perspective, an effective theory description of the initial state provides a reasonable framework in which to address other outstanding problems of inflation, such as the ``backreaction problem'' \cite{schalm2,backreact}, as well.

\begin{acknowledgments}

\noindent
This work was supported in part by DOE grant DE-FG03-91-ER40682 and the National Science Foundation grant PHY02-44801.  RH would like to thank Gary Shiu, Emil Mottola and Ira Rothstein for useful discussions, as well as the Michigan Center for Theoretical Physics for hospitality during the beginning of this work.

\end{acknowledgments}

\appendix

\section{The Schwinger-Keldysh Formalism}
\label{SKformal}

The Schwinger-Keldysh formalism \cite{schwinger,keldysh,kt} was developed to study the evolution of matrix elements over finite intervals of time.  It is therefore ideally suited not only to backgrounds which do not admit an $S$-matrix description, such as de Sitter space \cite{witten}, but also to settings where the state is specified at a particular time rather than in terms of its asymptotic properties.  In a standard $S$-matrix calculation, the goal is to determine the amplitude for a state in the far past, $|\psi\rangle$, to become some state $|\psi'\rangle$ in the far future,
\begin{equation}
\langle\psi'|S|\psi\rangle = \langle \psi'(\infty) | \psi(-\infty) \rangle . 
\label{Smatrixdef}
\end{equation}
Here, the states are assumed to be noninteracting in the asymptotic past and future and are usually taken to be the eigenstates of the free Hamiltonian, $H_0$, which are assumed to be in one-to-one correspondence with the eigenstates of the interacting theory as we adiabatically turn on and then turn off the interactions between $t=-\infty$ and $t=\infty$.  The scattering operator $S$ corresponds to the time-evolution operator evaluated over an infinite interval, 
\begin{equation}
S = U(\infty,-\infty) ,
\label{SisU}
\end{equation}
which is for a finite interval in the interaction picture, 
\begin{equation}
U(t,t') = Te^{-i\int_{t'}^t dt^{\prime\prime}\, H_I(t^{\prime\prime}) } , 
\label{Udyson}
\end{equation}
where $H_I(t)$ is the interacting part of the Hamiltonian.

The physical situation we are considering is quite different.  Instead of fixing the asymptotic properties of the state in the far past, the state is specified at a particular time, $t=t_0$,
\begin{equation}
|\psi(t_0)\rangle . 
\label{initstates}
\end{equation}
The state can be quite general and is not necessarily an eigenstate of either the free or the interacting theory.  For example, in an expanding background no true vacuum choice may exist and a particular state is chosen based upon its symmetries or as the result of some preceding dynamical model.  In this setting, the short distance properties of the state become more important since the cosmological expansion eventually redshifts these features to potentially observable scales.  Fixing an initial state allows for a more careful treatment of such effects.  

The set of measurable quantities is formed by the expectation values of operators, ${\cal O}(t)$, evaluated at some later time, $t>t_0$, rather than the overlap of past and future eigenstates of the free theory.  Therefore, it is more useful to evolve both the ``in'' and the ``out'' states forward, 
\begin{equation}
\langle \psi'(t) | {\cal O}_I(t) | \psi(t) \rangle . 
\label{SKmatrix}
\end{equation}
Here we have written the matrix element with two different states for generality although most often we shall set $\langle\psi'(t)| \to \langle\psi(t)|$.  In the interaction picture, the time-evolution operator of Eq.~(\ref{Udyson}) relates the future states to those we have specified at $t=t_0$, 
\begin{eqnarray}
&&\langle \psi'(t) | {\cal O}_I(t) | \psi(t) \rangle 
\nonumber \\
&&\qquad 
= \langle \psi'(t_0) | U^\dagger(t,t_0) {\cal O}_I(t) U(t,t_0) | \psi(t_0) \rangle . \qquad
\label{SKmatrixE}
\end{eqnarray}
The operator evolves too, but its evolution is determined by the free part of the Hamiltonian so that, in particular, if ${\cal O}_I$ corresponds to a product of fields, its time-evolution is already encoded in the time-dependence of the mode functions as in Eq.~(\ref{genmodes}).  Unlike the $S$-matrix approach above, the Schwinger-Keldysh approach contains two insertions of the time-evolution operator.

\begin{figure}[!tbp]
\includegraphics{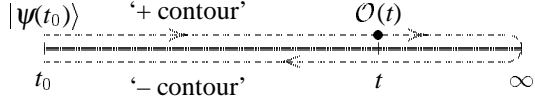}
\caption{The time contour---or equivalently the field content---can be formally doubled in order to write the time evolution of the initial and the final states, both defined at $t=t_0$, in terms of a single time evolution operator.
\label{ctp}}
\end{figure}

The time-dependent matrix element is usually written in a more compact form by formally doubling the time contour as follows.  We first insert a factor of the identity operator in the form $U^\dagger(\infty,t) U(\infty,t)$ to extend the time evolution to $t=\infty$
\begin{eqnarray}
&&\!\!\!\!\!\!\!\!\!\!\!\!\!\!\!
\langle \psi'(t) | {\cal O}_I(t) | \psi(t) \rangle 
\label{SKmatrixII} \\
&=& \langle \psi'(t_0) | U^\dagger(\infty,t_0) U(\infty,t) {\cal O}_I(t) U(t,t_0) | \psi(t_0) \rangle . 
\nonumber
\end{eqnarray}
Now double the time path so that times associated with the right three operators are formally considered to be on the ``$+$ contour,''
\begin{equation}
U(\infty,t) {\cal O}_I(t) U(t,t_0)
\to U(\infty,t^+) {\cal O}_I(t^+) U(t^+,t^+_0)
\label{Pcontour}
\end{equation}
and the times associated with the remaining operator are on the ``$-$ contour,''
\begin{equation}
U^\dagger(\infty,t_0) = U(t_0,\infty) \to U(t^-_0,\infty) . 
\label{Mcontour}
\end{equation}
Because of the Hermitian conjugation, the ``$-$ contour'' runs over the interval $[\infty,t_0^-]$ with the opposite time-ordering.  The points at $t^{\pm}=\infty$ are identified, so that we obtain a single contour running from $t=t_0^+$ to $t=\infty$ back to $t=t_0^-$ as shown in Fig.~\ref{ctp}.  
By using a time-ordering operator along this path, each piece of the time evolution is grouped into a single operator, 
\begin{eqnarray}
&&\!\!\!\!\!\!\!\!\!\!\!\!\!\!\!
\langle \psi'(t) | {\cal O}_I(t) | \psi(t) \rangle 
\label{SKmatrixIII} \\
&=& \langle \psi'(t_0) | T \bigl( {\cal O}_I(t) 
U(t_0^-,\infty) U(\infty,t_0^+) \bigr) | \psi(t_0) \rangle , 
\nonumber
\end{eqnarray}
which, using Eq.~(\ref{Udyson}), is given by 
\begin{eqnarray}
&&\!\!\!\!\!\!\!\!\!\!\!\!\!\!\!\!\!\!\!\!\!
U(t_0^-,\infty) U(\infty,t_0^+) 
\nonumber \\
&=& Te^{-i\int_\infty^{t_0^-} dt^{\prime\prime}\, H_I(t^{\prime\prime}) }
Te^{-i\int_{t_0^+}^\infty dt^{\prime\prime}\, H_I(t^{\prime\prime}) }
\nonumber \\
&=& 
Te^{-i\int_{t_0^+}^\infty dt^{\prime\prime}\, H_I(t^{\prime\prime}) 
+ i\int_{t_0^-}^\infty dt^{\prime\prime}\, H_I(t^{\prime\prime})}
\label{SKmatrixIV} 
\end{eqnarray}
Finally, let us replace the two parts of the path with a single interval, $t^{\prime\prime}\in [t_0,\infty]$, by equivalently doubling the field content of the theory.  Fields evaluated along the $\pm$ parts of the contour are written as $\varphi(t^\pm)=\varphi^\pm(t)$, and thus 
\begin{equation}
U(t_0^-,\infty) U(\infty,t_0^+) =
Te^{-i\int_{t_0}^\infty dt^{\prime\prime}\, \left[ H_I[\varphi^+(t^{\prime\prime})] - H_I[\varphi^-(t^{\prime\prime})] 
\right] } . 
\label{SKmatrixV} 
\end{equation}
The time-ordering of the fields inherit the time-ordering of the parts of the path on which they were originally defined, as will be explained in more detail below.  Inserting this expression into Eq.~(\ref{SKmatrixIII}) we obtain the general time-evolution of a matrix element
\begin{eqnarray}
&&\!\!\!\!\!\!\!\!\!\!\!\!\!\!\!
\langle \psi'(t) | {\cal O}_I(t) | \psi(t) \rangle 
\label{SKmatrixVI} \\
&=& \langle \psi'(t_0) | T \bigl( {\cal O}_I(t) 
e^{-i\int_{t_0}^\infty dt^{\prime\prime}\, 
\left[ H_I[\varphi^+] - H_I[\varphi^-] \right] } 
 \bigr) | \psi(t_0) \rangle .
\nonumber
\end{eqnarray}
Finally, we can remove the `vacuum-to-vacuum' portion of the matrix element and evaluate 
\begin{eqnarray}
&&\!\!\!\!\!\!\!\!\!\!\!\!\!\!\!
{\langle \psi'(t) | {\cal O}_I(t) | \psi(t) \rangle \over
\langle \psi'(t) | \psi(t) \rangle }
\label{SKevolve} \\
&=& {\langle \psi'(t_0) | T \bigl( {\cal O}_I(t) 
e^{-i\int_{t_0}^\infty dt^{\prime\prime}\, 
\left[ H_I[\varphi^+] - H_I[\varphi^-] \right] } 
 \bigr) | \psi(t_0) \rangle \over
\langle \psi'(t_0) | T \bigl( 
e^{-i\int_{t_0}^\infty dt^{\prime\prime}\, 
\left[ H_I[\varphi^+] - H_I[\varphi^-] \right] } 
 \bigr) | \psi(t_0) \rangle} . 
\nonumber
\end{eqnarray}
For example, setting $|\psi(t_0)\rangle = |\psi'(t_0)\rangle = |\alpha_k\rangle$ and ${\cal O}_I(t) =  \psi^+(x)$ we obtain Eq.~(\ref{nolinfull}).

The perturbative calculation of a matrix element now proceed similarly to that of the $S$-matrix element except that here we can encounter four different Wick contractions, 
\begin{eqnarray}
-iG_F^{\pm\pm}(x,y) 
&=& -i \int {d^3\vec k\over (2\pi)^3}\, e^{i\vec k\cdot (\vec x-\vec y)} G_k^{\pm\pm}(t,t')
\nonumber \\
&=& \langle\psi'(t_0) | T_{\alpha_k} \bigl( \varphi^\pm(x) \varphi^\pm(y) \bigr) | \psi(t_0) \rangle . \qquad
\label{pmGreens}
\end{eqnarray}
In our case, the initial and final states are the same and correspond to that which satisfies the linear initial condition given in Eq.~(\ref{BDflat}).  In terms of the vacuum Wightman functions,
\begin{equation}
G_k^>(t,t') = G_k^<(t',t) = i U_k^E(t) U_k^{E*}(t')
= {i\over 2\omega_k} e^{-i\omega_k(t-t')} ,
\label{awight}
\end{equation}
in the region $t>t_0$ the four propagators generalizing that which is consistent with the initial conditions, as given in Eq.~(\ref{dpropTHOI}), are  
\begin{eqnarray}
G_k^{++}(t,t') 
&\!=\!& \Theta(t-t') G_k^>(t,t') + \Theta(t'-t) G_k^<(t,t')
\nonumber \\
&& 
+ e^{\alpha_k} G_k^<(t_{\scriptscriptstyle I},t')
\nonumber \\
G_k^{--}(t,t') 
&\!=\!& \Theta(t'-t) G_k^>(t,t') + \Theta(t-t') G_k^<(t,t')
\nonumber \\
&& 
+ e^{\alpha_k} G_k^<(t_{\scriptscriptstyle I},t')
\nonumber \\
G_k^{-+}(t,t') 
&\!=\!& G_k^>(t,t') + e^{\alpha_k} G_k^<(t_{\scriptscriptstyle I},t')
\nonumber \\
G_k^{+-}(t,t') 
&\!=\!& G_k^<(t,t') + e^{\alpha_k} G_k^<(t_{\scriptscriptstyle I},t') . 
\label{pmGreensdet} 
\end{eqnarray}
Notice that because $t$ and $t'$ always occur after $t_0$, the image parts of these propagators, that proportional to $e^{\alpha_k}$, are all the same.

\section{Resummation of Logarithmic Divergences}
\label{logresum}

In arriving at the renormalization conditions of Sec.~\ref{rencon}, we made use of the so-called amplitude expansion approximation, i.e.\ the condition that $m^2\gg\lambda\phi^2$. This allows us to solve the mode equation in terms of simple plane waves $e^{-i\omega_k t}$ with $\omega_k = \sqrt{k^2+m^2}$. However, we can improve on this result by resumming insertions of $\lambda \phi_0^2$ into our Green's function. To do this, start with Eq.~(\ref{Hint}) and make the replacement
\begin{equation}
\label{eq:HintRepl}
{\textstyle{1\over 4}} \lambda \phi^2{\psi^\pm}^2 
\to {\textstyle{1\over 4}} \lambda \phi_0^2 {\psi^\pm}^2 
+ {\textstyle{1\over 4}} \lambda (\phi^2-\phi_0^2) {\psi^\pm}^2.
\end{equation}
We absorb the term ${\lambda\over 4} \phi_0^2{\psi^\pm}^2$ into a redefinition of the frequency of the modes
\begin{equation}
\label{eq: newfreq}
\omega_k \rightarrow \omega_{k0} 
\equiv \sqrt{k^2 + \left( m^2 + {\textstyle{1\over 2}} \lambda \phi_0^2 \right) } 
\equiv \sqrt{k^2 + {\cal M}_0^2}.
\end{equation}
This modification has interesting implications for the logarithmically divergent term in Eq.~(\ref{ibpdebris}) which was obtained by using integration by parts on
\begin{equation}
- {\lambda^2\over 8} \phi(t) \int_{t_0}^t dt'\, \phi^2(t')
\int {d^3\vec k\over (2\pi)^3}\, {\sin(2\omega_k(t-t'))\over\omega_k^2} . 
\end{equation}
Our redefinition of the frequencies implies that this is changed to 
\begin{equation}
- {\lambda^2\over 8} \phi(t) \int_{t_0}^t dt'\, 
\left( \phi^2(t') - \phi_0^2 \right)
\int {d^3\vec k\over (2\pi)^3}\, {\sin(2\omega_{k 0}(t-t'))\over\omega_{k 0}^2} . 
\label{eq:modcouplecor}
\end{equation}
Integrating the part containing $\phi^2(t^{\prime})$ by parts as before, yields
\begin{eqnarray}
&&\!\!\!\!\!\!\!\!\!\!\!\!\!\!
- {\lambda^2\over 8} \phi(t) \int_{t_0}^t dt'\, 
\left( \phi^2(t') - \phi_0^2 \right)
\int {d^3\vec k\over (2\pi)^3}\, {\sin(2\omega_{k 0}(t-t'))\over\omega_{k 0}^2} 
\nonumber \\
&=& 
- {\lambda^2\over 16} \phi^3(t) K(0) 
+ {\lambda^2\over 16} \phi(t) \phi^2(t_0) K(t-t_0) 
\nonumber \\
&&
+ {\lambda^2\over 8} \phi(t) \int_{t_0}^t dt'\, \phi(t') \dot\phi(t') K(t-t'),
\label{eq:modcouplibp}
\end{eqnarray}
where the kernel $K(t-t')$ is as given in Eq.~(\ref{kerneldef}), but with in terms of the new frequency $\omega_{k 0}$.  We see that the piece in Eq.~(\ref{eq:modcouplecor}) containing $\phi_0^2$ {\it exactly\/} cancels the term ${1\over 16} \lambda^2 \phi(t) \phi^2(t_0) K(t-t_0)$ in Eq.~(\ref{eq:modcouplibp}).  Thus the effect of this logarithmic divergence was absorbed into a renormalization of the frequency. 

There are other repercussions of this resummation.  Consider the term
\begin{eqnarray}
&&{i\lambda_R^2\over 8} \phi(t) \int_{t_0}^t dt'\, 
\left( \phi^2(t')-\phi^2_0 \right) 
\nonumber \\
&&\quad\times
\int {d^3\vec k\over (2\pi)^3}\, {e^{\alpha_k}\over\omega_{k 0}^2} 
\left[ e^{-2i\omega_{k 0}(t-t_0)} - e^{-2i\omega_{k 0}(t'-t_0)} \right] 
\qquad
\end{eqnarray}
If we make the same smoothness assumption about $\phi(t)$ as we made in Sec.~\ref{bndren}, we find that the integrand of the momentum integral is at least ${\cal O}\left( (t-t_0)^3 \right)$ and so will not contribute in the $t \to t_0$ limit. 

The rest of the calculation is as in the main text, with the replacement $\ln(\mu^2/m^2) \to \ln(\mu^2/{\cal M}_0^2)$.

\end{document}